\documentclass[11pt,a4paper]{article}
\pdfoutput=1
\usepackage{jheppub}
\usepackage{amsmath}
\usepackage{amssymb}
\usepackage{pdfsync}
\usepackage{shuffle}
\usepackage{slashed}
\usepackage{soul}

\usepackage{tikz}
\usetikzlibrary{decorations.pathmorphing}
\usetikzlibrary{arrows.meta}

\def\I{\tiny \mbox{I}}
\def\II{\tiny \mbox{II}}

\def\del #1{\mbox{\st{$#1$}}}

\title{From Operator Product Expansion to Anomalous Dimensions}
\author[a]{Rijun Huang}
\author[b]{, Qingjun Jin}
\author[b]{and Yi Li}

\affiliation[a]{Institute of Theoretical Physics, School of Physics and Technology, Nanjing Normal University, \\ No.1 Wenyuan Road, Nanjing 210046, P.R.China}
\affiliation[b]{Graduate School of China Academy of Engineering Physics, \\ No. 10 Xibeiwang East Road, Haidian District, Beijing, 100193, P.R.China }

\emailAdd{huang@njnu.edu.cn}
\emailAdd{qjin@gscaep.ac.cn}
\emailAdd{yili@gscaep.ac.cn}

\abstract{We propose a new method for computing the renormalization functions, which is based on the ideas of operator product expansion and large momentum expansion. In this method, the renormalization $Z$-factors are determined by the ultraviolet finiteness of Wilson coefficients in the dimensional regularization scheme. The ultraviolet divergence is extracted solely from two-point integrals at the large momentum limit. We develop this method in scalar field theories and establish a general framework for computing anomalous dimensions of fields, mass, couplings and composite operators. In particular, it is applied to  the 6-dimensional cubic scalar theory and the 4-dimensional quartic scalar theory. We demonstrate this method by computing the anomalous dimension of the $\phi^Q$ operator in cubic theory up to four loops for arbitrary $Q$, which is in agreement with the known result in the large $N$ limit. The idea of computing anomalous dimensions from the operator production expansion is general and can be extended beyond scalar theories. This is demonstrated through examples of the Gross-Neveu-Yukawa model with generic operators.}


\begin{document}
\maketitle \flushbottom

\section{Introduction}
\label{sec:introduction}

In quantum field theory, the renormalization was originally developed as a method to remove ultraviolet (UV) divergences and avoid infinities when computing Feynman integrals. It plays a crucial role in defining effective theories that predict and match experimental observables. As in the early development of Quantum Electrodynamics (QED) in 1940s, it resolved the disagreement between the theoretical divergent electron self-energy correction and the experimental finite physical observables \cite{Tomonaga:1946zz,Bethe:1947id,Schwinger:1948iu,Dyson:1949bp}. A significant achievement of renormalization in particle physics is the explanation of asymptotic freedom in Quantum Chromodynamics (QCD). Asymptotic freedom is manifested by the decreasing coupling with increasing energy, which can be understood through the computation of the beta function. As is well-known, this idea is crucial for establishing QCD as the theory of strong interaction \cite{Gross:1973id,Politzer:1973fx}. Apart from its early interests in particle physics, the renormalization group soon advanced into statistical physics and found itself highly suitable for understanding problems of phase transitions and critical phenomena. For example, the critical index in the scalar $\phi^4$ theory with $O(N)$ symmetry \cite{Jin:2022nqq,Bednyakov:2022guj} can describe the critical phenomena of some systems such as superfluid $^{4}$He, ferromagnetism, etc. In the condensed matter physics, the renormalization has been applied to the understanding of the quantum liquids or phases with broken symmetries, such as superconducting states \cite{Shankar:2017zag}. We suggest the readers consult related specialists if interested.

Initiated from QED, the idea of renormalization was reinforced and formalized by several works in the following 1950s \cite{Stueckelberg:1951gg,StueckelbergdeBreidenbach:1952pwl,Gell-Mann:1954yli,Bogolyubov:1959bfo}, and eventually evolved into the concept of {\sl continuum renormalization group}. In the continuum picture, physical quantities are required to be independent of the renormalization conditions, and equal to the observed values. Through the efforts of Callan and Symanzik \cite{Callan:1970yg,Symanzik:1970rt}, the continuum renormalization group found its extensive applications in particle physics, such as computing the beta functions and the anomalous dimensions of the field and mass. A decade later, Kenneth Wilson's series works laid the foundation for the modern understanding of  renormalization \cite{PhysRevB.4.3174,PhysRevB.4.3184,Wilson:1983xri}, known today as the {\sl Wilsonian renormalization group}. In the Wilsonian picture, the coupling constants in a theory change in such a way that the observables remain the same when the UV cutoff $\Lambda$ is changing. Furthermore, the Wilsonian picture provides a reasonable viewpoint for understanding the physical meaning of non-renormalizable theories. That is, one can regard a non-renormalizable theory as an effective field theory describing physics at low energies. Then, Joseph Polchinski's work in the mid-1980s clarified an approach in which the Wilsonian renormalization group can be implemented via the path integral formalism \cite{Polchinski:1983gv}. This method induces a set of differential equations on the couplings known as the renormalization group equations, and it gradually becomes a fundamental tool in quantum field theory with broad applications.

Working out the renormalization group equations in practical computation poses a significant challenge as it essentially involves Feynman graph and Feynman integral computations. It inherits the same difficulties as most problems based on the Feynman graph method.
The complexity escalates rapidly with increasing loop orders and the number of external momenta. Not only does the number of graphs increase factorially, but performing integration also becomes an arduous mathematical task that involves an increasing types and number of special functions.
Along the development of renormalization, numerous methods have been proposed to enhance the computation ability and efficiency, enabling one to take on the higher loop ($L\ge4$) challenge.
For example, the $R^\ast$ operation method \cite{CHETYRKIN1984419,Larin:2002sc}, which utilizes the fact that the UV counterterm of $(L+1)$-loop Feynman integral can be expressed as the $L$-loop integral with massless propagators, has been widely used in calculating the renormalization functions. The beta function in QCD theory has been generalized to five-loop order by the $R^\ast$ operation method \cite{Baikov:2016tgj}, although it took around 15-20 workstations and more than seven months to complete. However, when applied to integrals with more than five loops and/or non-trivial numerators such as multiple tensor structures of loop momenta, the $R^\ast$ operation method can be inefficient because it involves complicated subtraction of UV and IR sub-divergences.
An alternative approach that circumvents the subtraction of sub-divergences is the massive bubble approach \cite{Misiak:1994zw,Chetyrkin:1997fm}, which reduces the computation of UV divergences to the evaluation of massive vacuum bubbles.
It has been applied to, for instance, the four-loop order renormalization of the Gross-Neveu model \cite{Gracey:2016mio}.
However, the master integrals of vacuum bubble integrals are more difficult to evaluate than that of the massless propagators. Currently, the high precision numerical values of complete five-loop master integrals are still unavailable \cite{Luthe:2015ngq}.
Another promising method for the higher loop challenge is the graphical function method \cite{Schnetz:2013hqa}, which evaluates Feynman integrals in coordinate space. It has been applied to the scalar $\phi^4$ theory, and obtained the beta functions and mass anomalous dimensions to seven loops, and the scalar self-energy to eight loops \cite{Schnetz:2022nsc}. Such computations can be performed by a normal desktop in a reasonable computation time.
However, although the graphical function method is well-suited for computing scalar integrals, it is only getting started for general Feynman integrals with numerators \cite{Schnetz:2024qqt}.
Despite their achievements, each of the aforementioned methods has its own limitations, and we need new methods in order to deal with higher loop integrals with complicated numerators.

In this paper, we proposed a new method for the computation of renormalization functions based on the ideas of operator product expansion (OPE) and large momentum expansion. The OPE states that the product of two nearby operators, $\widehat{\mathcal{O}}_{\I}$ and  $\widehat{\mathcal{O}}_{\II}$, can be expanded by a list of local operators $\widehat{\mathcal{O}}_{i}$,
\begin{equation}
\widehat{\mathcal{O}}_{\II}(x)\widehat{\mathcal{O}}_{\I}(y)\xrightarrow{x\to y} \sum_{i}C_{i}(x-y)\widehat{\mathcal{O}}_{i}(y)~.
\end{equation}
When considering the renormalization of the operator, using the large momentum expansion, the Wilson coefficients $C_{i}$ can be expressed by a combination of $Z$-factors of $\widehat{\mathcal{O}}_{\I}$, $\widehat{\mathcal{O}}_{\II}$ and $\widehat{\mathcal{O}}_{i}$, times a propagator-type integral\footnote{A propagator-type integral refers to an integral with two external legs. Without confusion, we will call it a two-point function in this paper.}.
Given that the $\epsilon$ expansion of the propagator-type integral is known, the combination of $Z$-factors can be determined by the UV finiteness of the Wilson coefficient. Consequently, the $Z$-factor of a complicated operator can be expressed by the $Z$-factor of a relatively simple one, and the $Z$-factor of a given operator can be completely determined by a proper chosen set of OPE coefficients. The anomalous dimensions are subsequently extracted from the corresponding $Z$-factors.
Similar procedures can also be applied to the computation of beta functions.
In this way, we have reduced the problem of computing the renormalization factors to the evaluation of the propagator-type integrals.

This paper is organized as follows. In \S \ref{sec:RG-method}, we start with a brief review of the operator product expansion and the large momentum expansion. Then we describe the general framework for computing $Z$-factors from the OPE for the correlation functions. While this method is in principle general, throughout the derivation we will use the operator $\phi^Q$ in cubic scalar theory to illustrate the details of this method. In \S\ref{sec:RG-example}, as a practical application, we use this method to compute the anomalous dimension of the $\phi^Q$ operator in 6-dimensional $O(N)$ cubic scalar theory up to four loops. This demonstrates the computational algorithm and the validation of the OPE method. In \S\ref{sec:OPE-general}, we describe the new method for general operators beyond scalar theories, and explain its generality for a broad range of situations including the operators with derivatives and the operator mixing. Examples of Gross-Neveu-Yukawa model have been provided for demonstration. Discussions are provided in \S\ref{sec:conclusion}. In the appendix, an example of extracting UV divergence from two-point functions is provided in \S\ref{appendix:sample_GF}, and the complete result of four-loop correction of the scaling dimension $\Delta_Q$ is given in \S\ref{appendix:4-loop-result}.

\section{A method for the computation of renormalization functions}
\label{sec:RG-method}

In this section, we will develop a general method for the computation of renormalization functions. This method relies on the ideas of OPE and the large momentum expansion. Ultimately, the computation will be reduced to the problem of computing the propagator-type integrals. The basic strategy for deriving this method can be outlined as follows. Firstly, the correlation function of the operators and fundamental fields is expanded as products of two-point functions and lower-point correlation functions through the large momentum expansion. Then, by arranging the expansion terms, we identify the quantity of two-point functions multiplied by a combination of $Z$-factors as the Wilson coefficient of the leading-order operator basis in OPE. Finally, the UV finiteness of the Wilson coefficient in dimensional regularization parameter $\epsilon$ ensures that the product of two-point functions and the combination of $Z$-factors is also UV finite. This sets up a system of algebraic equations for $Z$-factors by requiring the cancelation of $\epsilon$ poles, provided the $\epsilon$ expansion of two-point functions have been computed\footnote{Similar ideas of extracting $Z$-factors from OPE coefficients were mentioned in \cite{Marino:2024uco}.}. This introduces the major improvement from our method, that in order to solve $Z$-factors, we only need to evaluate the integrals of two-point functions, which are much simpler than the original correlation functions. By solving the algebraic equations, the $Z$-factors can be determined recursively.

We will elaborate on the above strategy in details in the following subsections. After a brief review of the OPE and large momentum expansion techniques, in \S\ref{subsec:LME-LO}, we discuss the leading order contributions of large momentum expansion for the correlation functions of operators and fundamental fields in 6-dimensional cubic or 4-dimensional quartic scalar theories. We introduce the {\sl minimal-cut prescription}\footnote{Note that throughout this paper we use the notation {\sl cut} to represent the splitting of Feynman graphs, which should not be misunderstood as the famous {\sl unitarity-cut} method.} to identify graphs contributing to the leading order terms of large momentum expansion. In \S\ref{subsec:OPE-and-LME} we re-examine OPE in the viewpoint of the large momentum expansion, and present consistent discussions on the leading order contributions. With the theoretical background clarified, in \S\ref{subsec:full-expansion} we provide a general proof on our strategy of computing $Z$-factors, which would serve as a fundamental framework for designing computation algorithms.

\subsection{The operator product expansion}
\label{subsec:OPE}

The product of two local operators exhibits asymptotic behavior when their coordinates tend to each other. Such behavior can be described by the OPE method with an expansion into a well-defined basis of local operators of a given theory. Since the product of operators is in general singular when they approach each other, one expects a singular behavior for expansion coefficients with respect to the difference of coordinates. The expansion basis would be the tensor products of fields, and possibly also rely on the classical scaling dimensions $\Delta$, {\sl etc.} Hence the OPE of two operators  $\widehat{\mathcal{O}}_{\II}(x), \widehat{\mathcal{O}}_{\I}(y)$ generally takes a schematic form as,
\begin{equation}
\widehat{\mathcal{O}}_{\II}(x)\widehat{\mathcal{O}}_{\I}(y)\xrightarrow{x\to y} \sum_{i,\Delta,n}C_{\mu_1\cdots\mu_n}^{\Delta,i}(x-y)\widehat{\mathcal{O}}_{\Delta,i}^{\mu_1\cdots\mu_n}(y)~,\label{eqn:ope-expansion}
\end{equation}
where $\widehat{\mathcal{O}}_{\Delta,i}^{\mu_1\cdots \mu_n}$ belongs to a complete set of expansion basis defined in coordinate $y$, and $C_{\mu_1\cdots\mu_n}^{\Delta,i}(x-y)$'s are Wilson coefficients depending only on the difference of coordinates $x'=x-y$. The Lorentz indices of Wilson coefficients can be carried by $x'^\mu$ or spacetime measure $\eta^{\mu\nu}$, while the latter can be contracted with the operators of the expansion basis to produce lower rank ones. In this case, what remains is the tensor structure of $x'^\mu$, and the Wilson coefficient in general behaves as
\begin{equation}
C_{\mu_1\cdots\mu_n}^{\Delta,i}(x')=c_n^{\Delta,i}\frac{x'^{\mu_1}\cdots x'^{\mu_n}}{(x'^2)^{\frac{\Delta_{\I}+\Delta_{\II}-\Delta+n}{2}}}~,\label{eqn:OPE-general-c}
\end{equation}
where the singular behavior is determined by the scaling dimensions $\Delta_{\I},\Delta_{\II}$ of the original operators $\widehat{\mathcal{O}}_{\I},\widehat{\mathcal{O}}_{\II}$, and the coefficient $c_n^{\Delta,i}$ is a function of the logarithms $\ln (x'^2)$. As expected, the powers and logarithms of $x'^2$ enter into the Wilson coefficients, showing a generic asymptotic behavior of quantum field theory.  Accordingly, we can require the operators to be symmetric traceless tensors, making them irreducible representation of $SO(1,D-1)$ group. Then the correlation functions of these operators are also symmetric and traceless, and we do not need to worry about the mixing among the operators of different tensor ranks.

Now let us consider the correlation function of both sides of eqn.(\ref{eqn:ope-expansion}) with  fundamental fields $\phi(y_i)$'s, and obtain the following relations among correlation functions,
\begin{equation}
\Big\langle\widehat{\mathcal{O}}_{\II}(x)\widehat{\mathcal{O}}_{\I}(y)\phi(y_1)\cdots \phi(y_m)\Big\rangle\xrightarrow{x\to y} \sum_{i,\Delta,n}C_{\mu_1\cdots\mu_n}^{\Delta,i}(x-y)\Big\langle\widehat{\mathcal{O}}_{\Delta,i}^{\mu_1\cdots\mu_n}(y)\phi(y_1)\cdots \phi(y_m)\Big\rangle~.\label{eqn:OPE-FF}
\end{equation}
The LHS of above equation is a $(m+2)$-point correlation function of two operators and $m$ fields, while the RHS consists of $(m+1)$-point correlation functions\footnote{If the operators are replaced by fundamental fields, the above OPE also describes expansion of correlation functions of off-shell fields.}. The Wilson coefficients can be determined by matching the correlation functions on both sides of the equation. If all operators are renormalized, the full correlation functions on both sides are UV finite (free of $\epsilon$ poles in dimensional regularization), so the Wilson coefficients are also UV finite.

The OPE is closely related to the large momentum expansion. To see this, let us Fourier transfer both sides of the original OPE to the momentum space, and get
\begin{equation}
{\widehat{\mathcal{O}}}_{\II}(q_1)\widehat{\mathcal{O}}_{\I}(q_2):=\int {\rm d}^4 x ~{\rm d}^4 y~e^{iq_1\cdot x}\left(e^{-iq_1\cdot y}e^{iq_1\cdot y}\right)e^{iq_2\cdot y}\sum_{i}{C}_{i}(x-y)\widehat{\mathcal{O}}_i(y)=\widetilde{C}(q_1)\widehat{\mathcal{O}}(q_1+q_2)~.\label{eqn:OPE-momentum-space}
\end{equation}
The limit $x\to y$ in coordinate space corresponds to $q_1\to \infty$ in momentum space, corresponding to a large off-shell field, while $(q_1+q_2)$ remains finite. Thus we can take advantage of the large momentum expansion technique to study the asymptotic behaviors of correlation functions.

\subsection{The large momentum expansion}
\label{subsec:LME}

The large momentum expansion is a technique for evaluating the asymptotic expansion of Feynman integrals when two or more external momenta are large compared to the other scales \cite{Smirnov:2002pj}. For our purpose, we will focus on the large momentum expansion with two large external momenta, so that the integral contains a single hard scale. From the experience, we know that a Feynman integral $F_{\mathcal{G}}$ is expected to have the following behavior when expanded around small parameters such as, for instance, a ratio $m^2/q^2$, with $q$ being the large momentum, \cite{Smirnov:2002pj},
\begin{equation}
F_{\mathcal{G}}(q^2,m^2)\xrightarrow{{\rm large}~q^2~{\rm limit}}(q^2)^{\omega}\sum_{n=n_0}^{\infty}\sum_{j=0}^{2L}{c_{ni}} \left(\frac{m^2}{q^2}\right)^n \ln^j \frac{m^2}{q^2}~,
\end{equation}
where $L$ is the loop order of Feynman graph $\mathcal{G}$ and $\omega$ is the degree of divergence of the graph.
Although this expansion behavior is not a consequence of rigorous mathematical theorems, it is expected that such asymptotic expansion is a general property of Feynman integrals. The difference between the original Feynman integral and the asymptotic expansion to an order $n=N$ is of order ${O}((m^2/q^2)^{N})$, hence it is possible to take several expansion terms to ensure that the difference approaches to the infinitesimal. Furthermore, different expansion orders have different $q^2$-dependence, which enables us to distinguish contributions from different expansion series.

In a Feynman graph $\mathcal{G}$, the large momenta could have many ways of flowing in the sub-graph of $\mathcal{G}$. In each way, the graph $\mathcal{G}$ can be split into two graphs, one with large momentum flow, and the other with all soft momenta. Since the large momentum expansion is only applied in the large momentum flow, the asymptotic behavior of corresponding Feynman integral under large momentum expansion then takes the form \cite{Smirnov:2002pj}
\begin{equation}
F_{\mathcal{G}}\rightarrow \sum_{\gamma} F_{\mathcal{G}/\gamma} \circ\mathcal{M}_{\gamma}F_{\gamma}~,\label{eqn:LME}
\end{equation}
where $\gamma$ is a sub-graph of $\mathcal{G}$ with all large external and internal momenta, and we will call it {\bf hard graph}. The $\mathcal{M}_{\gamma}$ is an operation of Taylor expansion on the hard graph $\gamma$ with respect to the small expansion parameters. The ${\mathcal{G}/\gamma}$ is the graph of $\mathcal{G}$ by identifying all lines in $\gamma$ to a vertex, which effectively becomes an off-shell operator. We will call the graph ${\mathcal{G}/\gamma}$ as {\bf soft graph}. The summation is over all possible configurations of the hard graphs.

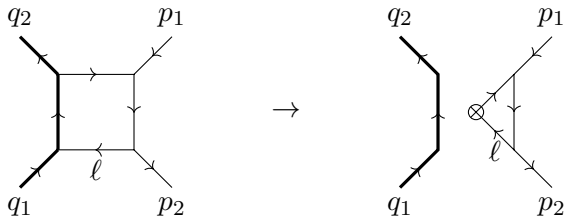
\begin{figure}
\centering
\begin{tikzpicture}
%
    \draw [very thick] (-1,-1)--(-0.5,-0.5) (-1,1)--(-0.5,0.5) (-0.5,-0.5)--(-0.5,0.5);
    \draw [] (-0.5,0.5)--(0.5,0.5)--(0.5,-0.5)--(-0.5,-0.5) (1,1)--(0.5,0.5) (1,-1)--(0.5,-0.5);
    \node [below] at (-1,-1) {$q_1$};
    \node [above] at (-1,1) {$q_2$};
    \node [above] at (1,1) {$p_1$};
    \node [below] at (1,-1) {$p_2$};
    \node [below] at (0,-0.5) {$\ell$};
    \draw [->] (0,-0.5)--(-0.01,-0.5);
    \draw [->] (-0.5,0)--(-0.5,0.01);
    \draw [->] (0,0.5)--(0.01,0.5);
    \draw [->] (0.5,0)--(0.5,-0.01);
    \draw [->] (-0.75,-0.75)--(-0.74,-0.74);
    \draw [->] (-0.75,0.75)--(-0.76,0.76);
    \draw [->] (0.75,-0.75)--(0.76,-0.76);
    \draw [->] (0.75,0.75)--(0.74,0.74);
    \node [] at (2.5,0) {$\rightarrow$};
    \draw [very thick] (4,-1)--(4.5,-0.5)--(4.5,0.5)--(4,1);
    \draw [] (5.5,0.5)--(5.5,-0.5)--(5,0)--(5.5,0.5) (6,1)--(5.5,0.5) (6,-1)--(5.5,-0.5);
    \draw [->] (4.5,-0.01)--(4.5,0);
    \draw [->] (5.5,0.01)--(5.5,0);
    \draw [->] (5.25,-0.25)--(5.24,-0.24);
    \draw [->] (5.25,0.25)--(5.26,0.26);
    \draw [->] (4.25,-0.75)--(4.26,-0.74);
    \draw [->] (4.25,0.75)--(4.24,0.76);
    \draw [->] (5.75,-0.75)--(5.76,-0.76);
    \draw [->] (5.75,0.75)--(5.74,0.74);
    \node [above] at (4,1) {$q_2$};
    \node [below] at (4,-1) {$q_1$};
    \node [above] at (6,1) {$p_1$};
    \node [below] at (6,-1) {$p_2$};
    \node [below] at (5.25,-0.25) {$\ell$};
    \draw [fill=white] (5, 0) circle [radius=0.1];
    \draw [] (4.93,-0.07)--(5.07,0.07) (4.93,0.07)--(5.07,-0.07);
\end{tikzpicture}
\caption{A scalar box integral at large momentum limit with large momentum flow represented by heavy lines, where external $q_1,q_2$ are taken to be large momenta. In large momentum expansion, this graph is effectively split into a hard graph and a soft graph. The former contains only large momentum lines, and the latter is produced by identifying all large momenta to a vertex while keeping soft momentum lines.}\label{fig:L1P4-example}
\end{figure}

The Taylor expansion $\mathcal{M}_\gamma$ defines the expansion series, and at the integrand level it is applied to the propagators with large momenta around small parameters. For example, if a propagator involves a large $q$ as $1/(\ell-q)^2$, in the small region of $\ell$, since both $|\ell^2|, |2q\cdot \ell|\ll q^2$, this propagator can be Taylor expanded around $q\to \infty$ as
\begin{equation}
\frac{1}{(\ell-q)^2}=\frac{1}{q^2}+\frac{2q\cdot \ell-\ell^2}{(q^2)^2}+\frac{(2q\cdot \ell-\ell^2)^2}{(q^2)^3}+\cdots
\end{equation}
The leading order contribution is simply given by setting the soft momentum $\ell$ to zero, which corresponds to replacing the propagator by $1/q^2$, while the higher order contributions contain polynomials of $\ell$ and $q$ in numerators. In the graph representation, the replacement of $1/q^2$ changes the corresponding internal lines, which effectively splits the original Feynman graph to a hard graph and a soft graph. This provides a pictorial description of eqn.(\ref{eqn:LME}). For example, a scalar box integral at the large momentum limit of two external large momenta $q_1,q_2$ with large momentum flow represented by heavy lines, can be expanded as a single line of large momenta times an 1-loop form factor, as sketched in Fig.(\ref{fig:L1P4-example}).
The leading order term of the integrand is then given by the product of hard graph and soft graph. For higher order terms of asymptotic series, there will be powers of $\ell$ in the numerators, and eventually we get expansion terms with tensor structures of $\ell$. The splitting of graph $\mathcal{G}$ generally separates the integration parameters, and makes the integration simpler than the original one. Although the tensor structures in the numerators would make things complicated, it does not trouble us since in this paper we will only use the information of the leading order contributions. It is sufficient to concentrate on the leading order terms for developing a method of computing renormalization functions.

\subsection{Determining leading order term of large momentum expansion}
\label{subsec:LME-LO}


The asymptotic expansion will express the integral as series expansion of $1/q^2$, where $q$ is the external momentum taken to the large limit. As mentioned, each expansion term has distinct $1/q^2$-dependence, so the contributions of different order terms will not mix together, and the leading order term is free of tensor structures.

The leading order term of an integral under the large momentum expansion is explicitly computed by the summation of products of hard and soft graphs, as described in eqn.(\ref{eqn:LME}). Pictorially, we define a terminology of {\sl cutting} to describe the splitting of Feynman graph. The hard and soft graphs are generated by cutting graph $\mathcal{G}$ at all soft momentum lines that connecting directly to the large momentum flow, and splitting the graph into hard sub-graph with only large momentum lines and the soft sub-graph with only soft momentum lines. For the latter one, all soft momentum lines that being cut are re-connected to a vertex\footnote{Pictorially, the soft graph can also be considered as shrinking all lines of the large momenta to a vertex.}. From the viewpoint of external momenta, we are in fact considering all possible cuttings that can separate the large external momenta with all other soft external momenta, in the condition that the hard graph are connected graph. Such cuttings produce all possible large momentum flow of a graph.

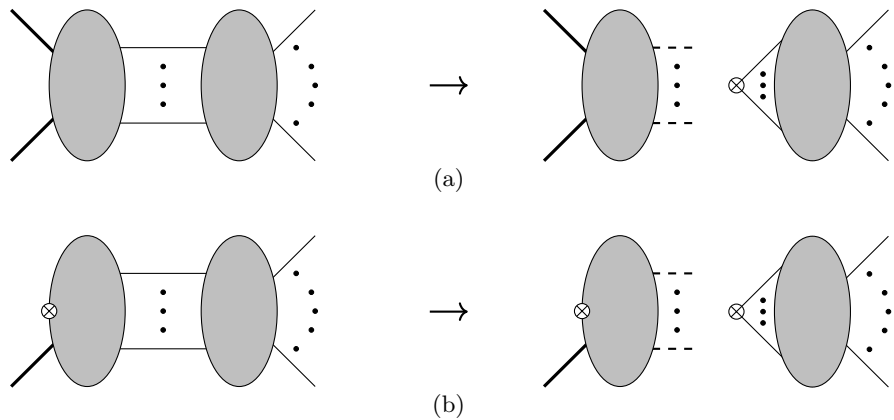
\begin{figure}
  \centering
    \begin{tikzpicture}
%
    \node [] at (2,0) {
    \begin{tikzpicture}
      \draw [very thick] (1,0)--(0,1) (1,0)--(0,-1);
      \draw [] (3,0)--(4,1) (3,0)--(4,-1);
      \draw [] (1,0.5)--(3,0.5) (1,-0.5)--(3,-0.5);
      \draw [fill=black] (2,0.25) circle [radius=0.03]  (2,0) circle [radius=0.03] (2,-0.25) circle [radius=0.03];
      \draw [fill=black] (3.75,0.5) circle [radius=0.03] (3.95,0.25) circle [radius=0.03] (4,0) circle [radius=0.03] (3.75,-0.5) circle [radius=0.03] (3.95,-0.25) circle [radius=0.03];
      \draw [fill=lightgray] (1,0) ellipse [x radius=0.5, y radius=1] (3,0) ellipse [x radius=0.5, y radius=1];
    \end{tikzpicture}
    };
    \node [] at (8,0) {
    \begin{tikzpicture}
      \draw [very thick] (1,0)--(0,1) (1,0)--(0,-1);
      \draw [dashed, thick] (1,0.5)--(2,0.5) (1,-0.5)--(2,-0.5);
      \draw [fill=black] (1.75,0.25) circle [radius=0.03] (1.75,0) circle [radius=0.03] (1.75,-0.25) circle [radius=0.03];
      \draw [fill=lightgray] (1,0) ellipse [x radius=0.5, y radius=1];
    \end{tikzpicture}
    };
    \node [] at (10.5,0) {
    \begin{tikzpicture}
      \draw [] (0,0)--(1,1) (0,0)--(1,-1);
      \draw [] (1,0)--(2,1) (1,0)--(2,-1);
      \draw [fill=black] (0.35,0.15) circle [radius=0.03] (0.35,0) circle [radius=0.03] (0.35,-0.15) circle [radius=0.03];
      \draw [fill=black] (1.75,0.5) circle [radius=0.03] (1.95,0.25) circle [radius=0.03] (2,0) circle [radius=0.03] (1.75,-0.5) circle [radius=0.03] (1.95,-0.25) circle [radius=0.03];
      \draw [fill=lightgray] (1,0) ellipse [x radius=0.5, y radius=1];
      \draw [fill=white] (0, 0) circle [radius=0.1];
      \draw [] (0.07,0.07)--(-0.07,-0.07) (-0.07,0.07)--(0.07,-0.07);
    \end{tikzpicture}
    };
    \draw [->,thick] (5.5,0)--(6,0);
    \node [] at (5.75,-1.25) {{\footnotesize (a)}};
    \node [] at (2,-3) {
    \begin{tikzpicture}
      \draw [very thick]  (1,0)--(0,-1);
      \draw [] (3,0)--(4,1) (3,0)--(4,-1);
      \draw [] (1,0.5)--(3,0.5) (1,-0.5)--(3,-0.5);
      \draw [fill=black] (2,0.25) circle [radius=0.03]  (2,0) circle [radius=0.03] (2,-0.25) circle [radius=0.03];
      \draw [fill=black] (3.75,0.5) circle [radius=0.03] (3.95,0.25) circle [radius=0.03] (4,0) circle [radius=0.03] (3.75,-0.5) circle [radius=0.03] (3.95,-0.25) circle [radius=0.03];
      \draw [fill=lightgray] (1,0) ellipse [x radius=0.5, y radius=1] (3,0) ellipse [x radius=0.5, y radius=1];
      \draw [fill=white] (0.5, 0) circle [radius=0.1];
      \draw [] (0.57,0.07)--(0.43,-0.07) (0.43,0.07)--(0.57,-0.07);
    \end{tikzpicture}
    };
    \node [] at (8,-3) {
    \begin{tikzpicture}
      \draw [very thick] (1,0)--(0,-1);
      \draw [dashed, thick] (1,0.5)--(2,0.5) (1,-0.5)--(2,-0.5);
      \draw [fill=black] (1.75,0.25) circle [radius=0.03] (1.75,0) circle [radius=0.03] (1.75,-0.25) circle [radius=0.03];
      \draw [fill=lightgray] (1,0) ellipse [x radius=0.5, y radius=1];
      \draw [fill=white] (0.5, 0) circle [radius=0.1];
      \draw [] (0.57,0.07)--(0.43,-0.07) (0.43,0.07)--(0.57,-0.07);
    \end{tikzpicture}
    };
    \node [] at (10.5,-3) {
    \begin{tikzpicture}
      \draw [] (0,0)--(1,1) (0,0)--(1,-1);
      \draw [] (1,0)--(2,1) (1,0)--(2,-1);
      \draw [fill=black] (0.35,0.15) circle [radius=0.03] (0.35,0) circle [radius=0.03] (0.35,-0.15) circle [radius=0.03];
      \draw [fill=black] (1.75,0.5) circle [radius=0.03] (1.95,0.25) circle [radius=0.03] (2,0) circle [radius=0.03] (1.75,-0.5) circle [radius=0.03] (1.95,-0.25) circle [radius=0.03];
      \draw [fill=lightgray] (1,0) ellipse [x radius=0.5, y radius=1];
      \draw [fill=white] (0, 0) circle [radius=0.1];
      \draw [] (0.07,0.07)--(-0.07,-0.07) (-0.07,0.07)--(0.07,-0.07);
    \end{tikzpicture}
    };
    \draw [->,thick] (5.5,-3)--(6,-3);
    \node [] at (5.75,-4.25) {{\footnotesize (b)}};
    \end{tikzpicture}
  \caption{Contributions of the large momentum expansion by cutting $k$ soft lines. The dots in the middle of graph represents $k$ small propagators or external legs to be cut. The dashed lines represent null momenta in momentum space or removed legs in coordinate space. (a) A $k$-cutting on the graph of correlation function of fundamental fields, producing the two-point functions of hard fields and form factors of soft fields. (b) A $k$-cutting on the graph of correlation functions of one operator inserted in fundamental fields, producing the two-point functions of a hard field and an off-shell operator, and form factors of soft fields. }\label{fig:k-cut}
\end{figure}

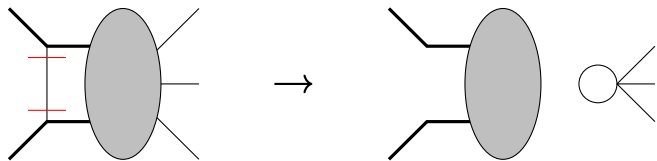
\begin{figure}
\centering
\begin{tikzpicture}
%
    \node [] at (1.25,0) {
    \begin{tikzpicture}
      \draw [very thick] (0,1)--(0.5,0.5)--(1.5,0.5) (0,-1)--(0.5,-0.5)--(1.5,-0.5);
      \draw [ ] (0.5,0.5)--(0.5,-0.5);
      \draw [] (1.5,0)--(2.5,1) (1.5,0)--(2.5,-1) (1.5,0)--(2.5,0);
      \draw [fill=lightgray] (1.5,0) ellipse [x radius=0.5,y radius=1];
      \draw [red] (0.25,0.35)--(0.75,0.35) (0.25,-0.35)--(0.75,-0.35);
    \end{tikzpicture}
    };
    \node [] at (6.75,0) {
    \begin{tikzpicture}
      \draw [very thick] (0,1)--(0.5,0.5)--(1.5,0.5) (0,-1)--(0.5,-0.5)--(1.5,-0.5);
      \draw [fill=lightgray] (1.5,0) ellipse [x radius=0.5,y radius=1];
      \draw [] (2.75,0) circle [radius=0.25];
      \draw [] (3,0)--(3.5,0) (3,0)--(3.5,0.5) (3,0)--(3.5,-0.5);
    \end{tikzpicture}
    };
    \draw [->,thick] (3.5,0)--(4,0);
\end{tikzpicture}
\caption{If cutting one internal line off the large momentum flow, we will get a tadpole soft graph, which is a scaleless integral in massless theory and should be excluded.}\label{fig:LME-tadpole}
\end{figure}

Applying the cutting prescription to the Feynman graphs of correlation functions with fundamental fields (or fields with an operator inserted), and setting the momenta of two external legs (or one external leg and one operator) to the large limit, the splitting of graphs can be represented as in Fig.(\ref{fig:k-cut}).
Note that the $k$-cutting generates a two-point hard graph and a soft graph of fields with an operator inserted. The latter will be evaluated as a form factor describing the off-shell operator to $k$ external fields. Note that the original graph describes a $(k+2)$-point correlation function (or a form factor of an operator to $(k+1)$ external fields), hence the $k$-cutting relates the original correlation function to the lower-point ones, with two-point functions as expansion coefficients. In general, all allowed cuttings applying on the internal and external lines that split a graph into the hard graph and the soft graph will contribute to the asymptotic expansions. There is no constraint on the number of lines being cut. This of course will produce a large amount of splitting possibilities for a graph. However we will soon show that, the lowest order contribution of $1/q^2$ that can be detected by a cutting depends on the number of lines being cut. By selecting cuttings of a specific $k$ lines, we can extract information for terms contributing to a certain order of $1/q^2$ in the asymptotic expansion. Before a detailed discussion, we should mention that there is one type of cutting that should be excluded in the consideration. If a propagator of the soft momentum is attached directly to the large momentum flow, after cutting it off and connecting two ends of the internal line to a vertex, we would get a tadpole soft graph, as shown in Fig.(\ref{fig:LME-tadpole}). In massless case it contributes to a scaleless integral and should be excluded. In the current paper, we are considering massless theories, so only the cuttings on different soft lines are considered.

In order to explicitly discuss the contributions of large momentum expansion in the prescription of $k$-cutting, from now onward we will consider the scalar theories with a so-called $\phi^Q$ operator\footnote{We postpone the explicit definition of $\phi^Q$ operator in \S\ref{sec:RG-example}.}. Now let us consider the $k$-cutting of a graph shown in Fig.(\ref{fig:k-cut}), and remark that the lines being cut could be the internal propagators or external soft legs. The large momentum expansion is determined by the hard graph, since it contains all large momenta. We want to determine the number of $k$ that contributing to the leading order terms of large momentum expansion. Let us suppose the hard graph contains $n_v$ vertices, $n_e$ propagators, and $L=n_e-n_v+1$ number of loops. The integral of hard graph scales in the large momentum as
\begin{equation}
\int \frac{{\rm{d}}^D\ell_1{\rm{d}}^D\ell_2\cdots {\rm{d}}^D\ell_L}{\mathcal{D}_1\mathcal{D}_2\cdots \mathcal{D}_{n_e}}~\sim \frac{1}{(q^2)^{-\frac{DL}{2}+n_e}}~,
\end{equation}
where $D$ is the spacetime dimension and $\mathcal{D}_i$'s are the propagators. Specifically, since we will consider the scalar theories in dimensional regularization scheme in $(4-2\epsilon)$ or $(6-2\epsilon)$ dimension, the scale behavior can be explicitly written as
\begin{equation}
D=4-2\epsilon~~:~~ \frac{1}{(q^2)^{2n_v-n_e-2+\epsilon L}}~~~,~~~D=6-2\epsilon~~:~~\frac{1}{(q^2)^{3n_v-2n_e-3+\epsilon L}}~.
\end{equation}
The $1/(q^2)^{\epsilon L}$ factor after expansion around the small $\epsilon$ provides the logarithmic dependence of $1/q^2$, while the other factor provides the series expansion of $1/q^2$.

For various theories and different correlation functions, by inspecting the structures of hard graphs, we can compute the relations of $n_v,n_e$ and the number $k$ of cuttings. These relations set constraints on the behaviors of $1/q^2$ expansion, and consequently the leading order contribution of large momentum expansion depends on the theories and physics quantities to be computed. If we are considering the two-point function of hard graph with only cubic vertices in $D=6-2\epsilon$ dimension, after cutting $k$ lines of the original graph, the hard graph has
$n_e$ internal lines and $n_v$ vertices. Then counting number of lines and vertices, we get
\begin{equation}
3n_v-(k+2)=2n_e~~\to ~~ 3n_v-2n_e=k+2~.\label{eqn:VE-condition-cubic-A}
\end{equation}
Similarly, if we consider the two-point function of a hard field and a $\phi^Q$ operator in above theory, the hard graph has one vertex of operator, $(n_v-1)$ cubic vertices
and $n_e$ internal lines. Then the counting leads to
\begin{equation}
  3(n_v-1)+Q-(k+1)=2n_e~~\to~~3n_v-2n_e=k+4-Q~.\label{eqn:VE-condition-cubic-FF}
\end{equation}
In both cases, we get the relations of scale behaviors and number of cuttings as
\begin{equation}
\begin{array}{l}
{\small\mbox{two-point~function}}\\
{\small\mbox{of~hard~fields}}
\end{array}
~:~~\frac{1}{(q^2)^{k-1+\epsilon L}}~~~,~~~
\begin{array}{l}
{\small\mbox{two-point~function}}\\
{\small\mbox{of~hard~field~and~operator}}
\end{array}~:~~\frac{1}{(q^2)^{k-(Q-1)+\epsilon L}}~.\label{eqn:cubic-q2-LO}
\end{equation}
The order of $1/q^2$ depends on the number $k$. The minimal number of cuttings is determined by the symmetry of theories. For the correlation function of fundamental fields in cubic theory, the minimal number of cuttings is $k=1$. While for the correlation function of fields with $\phi^Q$ operator inserted, by charge conservation the minimal number of cuttings is $k=Q-1$. So from eqn.(\ref{eqn:cubic-q2-LO}) we conclude that the leading order term is given by the minimal cuts. Cutting more lines produces higher order contributions of large momentum expansion.

The same discussion can be applied to the theories with quartic vertices in $D=4-2\epsilon$ dimension. In such a theory, counting number of lines and vertices yields
\begin{equation}
\begin{array}{l}
{\small\mbox{two-point~function}}\\
{\small \mbox{of~hard~fields}}
\end{array}~:~~2n_v-n_e=\frac{k}{2}+1~~~,~~~\begin{array}{l}
{\small\mbox{two-point~function}}\\
{\small\mbox{of~hard~field~and~operator}}
\end{array}~:~~2n_v-n_e=\frac{k-Q+5}{2}~.\label{eqn:VE-condition-quartic}
\end{equation}
Then the scale behavior of $1/q^2$ depends on the number of cuttings as
\begin{equation}
\begin{array}{l}
{\small\mbox{two-point~function}}\\
{\small \mbox{of~hard~fields}}
\end{array}~:~~\frac{1}{(q^2)^{\frac{k}{2}-1+\epsilon L}}~~~,~~~\begin{array}{l}
{\small\mbox{two-point~function}}\\
{\small\mbox{of~hard~field~and~operator}}
\end{array}~:~~\frac{1}{(q^2)^{\frac{k-(Q-1)}{2}+\epsilon L}}~.\label{eqn:quartic-q2-LO}
\end{equation}
By the symmetry consideration, the minimal number of cuttings for two-point function of fields in quartic theory is $k=2$. While by charge conservation consideration, the minimal number of cuttings for two-point function of fields with $\phi^Q$ operator inserted is $k=Q-1$. From eqn.(\ref{eqn:quartic-q2-LO}) we conclude again the leading order term is given by the minimal cuts of corresponding cases.

Computing the $k$-cuttings of the Feynman graph rather than the minimal cuts, we will get the higher order terms of the $1/q^2$ expansion. The Taylor expansion of propagators will also generate higher order terms of $1/q^2$ with tensor structures in the numerator. The mixing origins of higher order terms makes it a bit complicated to compute contributions beyond leading order. However, since each order of $1/q^2$ expansion is independent, we can restrict ourselves within the leading order contributions produced by the minimal cuts of graph. In paper \cite{Schnetz:2016fhy,Schnetz:2022nsc}, similar asymptotic expansion of graph is discussed, which is in fact the large momentum expansion in coordinate space. In their treatment, all possible number of cuts are considered, leading to the results involving higher order contributions. Our treatment of minimal cuts greatly reduces the number of expansion terms, and simplifies the computation.

The Feynman integrals can be evaluated in the momentum space or the coordinate space, depending on the problems considered and the computational tools available. Nevertheless, we can also analyze the scale behaviors of the hard graph in coordinate space, to get the leading order dependence of $1/q^2$ expansion under the $k$-cutting. In coordinate space, the large momentum limit is realized by the limit $x_Q\to 0$, where $x_Q$ is the coordinate of large leg. The asymptotic expansion is then described by series expansion of $x_Q^2$. The scale behavior of two-point function of fields is slightly different from that of fields with $\phi^Q$ operator inserted, since for the latter, the vertex of operator is considered to be external vertex and should not be integrated. Following the hard graph in Fig.(\ref{fig:k-cut}), we can compute the scale behavior of $x_Q^2$ as
\begin{eqnarray}
&&\begin{array}{l}
{\small\mbox{two-point~function}}\\
{\small \mbox{of~hard~fields}}
\end{array}~:~~\int \frac{{\rm d}^D y_1{\rm d}^D y_2\cdots {\rm d}^D y_{n_v}}{\Big(X_1^2\cdots X_{n_e}^2(y_i-x_Q)^2(y_j-x'^2_Q)\Big)^{\lambda}}\sim {(x_Q^2)^{\frac{D n_v}{2}-\lambda(n_e+2)}}~,\\
&&\begin{array}{l}
{\small\mbox{two-point~function}}\\
{\small\mbox{of~hard~field~and~operator}}
\end{array}~:~~\int \frac{{\rm d}^D y_1{\rm d}^D y_2\cdots {\rm d}^D y_{n_v-1}}{\Big(X_1^2\cdots X_{n_e}^2(y_i-x_Q)^2\Big)^{\lambda}}\sim {(x_Q^2)^{\frac{D (n_v-1)}{2}-\lambda(n_e+1)}}~,
\end{eqnarray}
where $\lambda=\frac{D-2}{2}$, the $y_i$'s are labels of internal vertices, the $x_Q,x'_Q$ are labels of external legs, while the $X_i^2$'s are propagators in coordinate space. We still have relation $n_e-n_v=L-1$, and eqn.(\ref{eqn:VE-condition-cubic-A}), eqn.(\ref{eqn:VE-condition-cubic-FF}), eqn.(\ref{eqn:VE-condition-quartic}) for cubic and quartic theories. With these relations, we get the scale behaviors of hard graphs as,
\begin{center}
  \begin{tabular}{|l|c|c|}
    \hline
    ~ & $D=6-2\epsilon$, cubic theory &  $D=4-2\epsilon$, quartic theory  \\ \hline
    $\begin{array}{l}
{\small\mbox{two-point~function}}\\
{\small \mbox{of~hard~fields}}
\end{array}$ & $(x_Q^2)^{k-2+\epsilon(L+1)}$ & $(x_Q^2)^{\frac{k}{2}-1+\epsilon(L+1)}$ \\ \hline
    $\begin{array}{l}
{\small\mbox{two-point~function}}\\
{\small\mbox{of~hard~field~and~operator}}
\end{array}$ & $(x_Q^2)^{k-(Q-1)-2+\epsilon(L+1)}$ & $(x_Q^2)^{\frac{k-(Q-1)}{2}-1+\epsilon(L+1)}$ \\
    \hline
  \end{tabular}
\end{center}
Similar as discussed in the momentum space, the leading order term of $x_Q^2$ expansion is determined by the minimal cuts. For cases with the $\phi^Q$ operator, the minimal number is $k=Q-1$, while for the two-point function of fundamental fields in cubic theory, the minimal number is $k=1$, and in quartic theory the minimal number is $k=2$. So the leading order term of hard graphs scales as,
\begin{center}
  \begin{tabular}{|l|c|c|}
    \hline
    ~ & $D=6-2\epsilon$, cubic theory &  $D=4-2\epsilon$, quartic theory  \\ \hline
    $\begin{array}{l}
{\small\mbox{two-point~function}}\\
{\small \mbox{of~hard~fields}}
\end{array}$ &  $\frac{1}{x_Q^2}$ & 1 \\ \hline
    $\begin{array}{l}
{\small\mbox{two-point~function}}\\
{\small\mbox{of~hard~field~and~operator}}
\end{array}$ &  $\frac{1}{(x_Q^2)^2}$ &  $\frac{1}{x_Q^2}$ \\
    \hline
  \end{tabular}
\end{center}

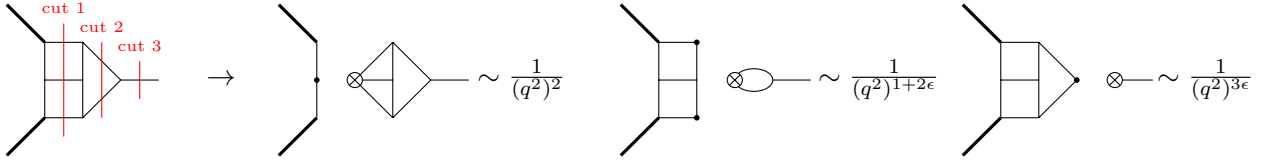
\begin{figure}
  \centering
    \begin{tikzpicture}
%
    \node [] at (1,0) {
    \begin{tikzpicture}
      \draw [very thick] (0.5,0.5)--(0,1) (0.5,-0.5)--(0,-1);
      \draw [] (0.5,0.5)--(0.5,-0.5) (0.5,0.5)--(1,0.5) (0.5,0)--(1,0) (0.5,-0.5)--(1,-0.5) (1,0.5)--(1,-0.5) (1,0.5)--(1.5,0) (1,-0.5)--(1.5,0) (1.5,0)--(2,0);
      \draw [red] (0.75,0.75)--(0.75,-0.75) (1.25,0.5)--(1.25,-0.5) (1.75,0.25)--(1.75,-0.25);
      \node [above,red] at (0.75,0.75) {{\tiny cut 1}};
      \node [below,white] at (0.75,-0.75) {{\tiny cut 1}};
      \node [above,red] at (1.25,0.5) {{\tiny cut 2}};
      \node [above,red] at (1.75,0.25) {{\tiny cut 3}};
    \end{tikzpicture}
    };
    \node [] at (4.75,0) {
    \begin{tikzpicture}
      \draw [very thick] (0.5,0.5)--(0,1) (0.5,-0.5)--(0,-1);
      \draw [] (0.5,0.5)--(0.5,-0.5) (1,0)--(1.5,0.5) (1,0)--(1.5,0) (1,0)--(1.5,-0.5) (1.5,0.5)--(1.5,-0.5) (1.5,0.5)--(2,0) (1.5,-0.5)--(2,0) (2,0)--(2.5,0);
      \draw [fill=black] (0.5,0) circle [radius=0.03];
      \draw [fill=white] (1, 0) circle [radius=0.1];
      \draw [] (1.07,0.07)--(0.93,-0.07) (0.93,0.07)--(1.07,-0.07);
    \end{tikzpicture}
    };
    \node [] at (9.25,0) {
    \begin{tikzpicture}
      \draw [very thick] (0.5,0.5)--(0,1) (0.5,-0.5)--(0,-1);
      \draw [] (0.5,0.5)--(0.5,-0.5) (0.5,0.5)--(1,0.5) (0.5,0)--(1,0) (0.5,-0.5)--(1,-0.5) (1,0.5)--(1,-0.5) (2,0)--(2.5,0);
      \draw [] (1.75,0) ellipse [x radius=0.25, y radius=0.15];
      \draw [fill=black] (1,0.5) circle [radius=0.03] (1,-0.5) circle [radius=0.03];
      \draw [fill=white] (1.5, 0) circle [radius=0.1];
      \draw [] (1.57,0.07)--(1.43,-0.07) (1.43,0.07)--(1.57,-0.07);
    \end{tikzpicture}
    };
    \node [] at (13.75,0) {
    \begin{tikzpicture}
      \draw [very thick] (0.5,0.5)--(0,1) (0.5,-0.5)--(0,-1);
      \draw [] (0.5,0.5)--(0.5,-0.5) (0.5,0.5)--(1,0.5) (0.5,0)--(1,0) (0.5,-0.5)--(1,-0.5) (1,0.5)--(1,-0.5) (1,0.5)--(1.5,0) (1,-0.5)--(1.5,0) (2,0)--(2.5,0);
      \draw [fill=black] (1.5,0) circle [radius=0.03];
      \draw [fill=white] (2, 0) circle [radius=0.1];
      \draw [] (2.07,0.07)--(1.93,-0.07) (1.93,0.07)--(2.07,-0.07);
    \end{tikzpicture}
    };
    \node [] at (2.75,0) {$\to$};
    \node [] at (6.7,0) {$\sim\frac{1}{(q^2)^{2}}$};
    \node [] at (11.4,0) {$\sim\frac{1}{(q^2)^{1+2\epsilon}}$};
    \node [] at (15.7,0) {$\sim\frac{1}{(q^2)^{3\epsilon}}$};
    \end{tikzpicture}
  \caption{Examples of the $k$-cuttings of a Feynman graph for scalar cubic theory in $D=6-2\epsilon$ dimension. Different number of cutting contributes to different orders of $1/q^2$ expansion.}\label{fig:cubic-expansion}
\end{figure}

In our method of computing renormalization functions, we will make use of the information of leading order contribution under the large momentum expansion. From above discussion we know that, it can be extracted by the minimal cuts of Feynman graph either in the momentum or coordinate space. Fig.(\ref{fig:cubic-expansion}) is an example of the scalar cubic theory in $D=6-2\epsilon$ dimension showing the scale behaviors infected by the number of cuttings, where the leading order term is given by $k=1$ cutting shown in the third graph. Notice that the contributions of the $k=2$ cutting and $k=3$ cutting both contain scaleless tadpole integrals in the soft graphs, thus they will not contribute to the final result. This observation makes it trivial for the asymptotic expansion of the three-point functions with two large external momenta in cubic theory, since the only possible way of splitting the hard and soft graphs without scaleless integrals is the $k=1$ cutting that removing the soft external leg from the original graph, leaving only two-point functions.
%

\subsection{Analyzing the leading order term from OPE}
\label{subsec:OPE-and-LME}

The OPE in momentum space eqn.(\ref{eqn:OPE-momentum-space}) clearly shows that the asymptotic expansion at large momentum limit is described by the series expansion of operator basis defined in the soft region, while the scale behavior of large momentum is encoded in the Wilson coefficients. Acting OPE on fundamental fields as eqn.(\ref{eqn:OPE-FF}), we get an expansion of the correlation functions into lower-point ones. Compared to the large momentum expansion as represented in Fig.(\ref{fig:k-cut}), we can infer that the Wilson coefficient corresponds to the two-point hard graph. An explicit connection between the Wilson coefficients and the hard graphs will be presented in the next subsection. In our setup, we consider the asymptotic expansion of two large external momenta, which corresponds to the OPE of two operators. Then the Wilson coefficients of OPE can be computed by the hard graphs of large momentum expansion, and the correlation function of operator basis with fundamental fields is equivalent to the soft graphs. Different cuttings in large momentum expansion contribute to the Wilson coefficients of different operator basis as well as different orders of asymptotic expansion over $x_Q^2$ in coordinate space. A complete operator basis contains tensor operators, which can be identified as the higher order terms of Taylor expansion in large momentum expansion. However, at the leading order, we only need to focus on the lowest order of operator basis.

The leading order term of asymptotic expansion in the large momentum expansion can also be analyzed by the OPE, by counting the mass dimension of Wilson coefficients. From now on let us again focus on the scalar theories with the $\phi^Q$ operator for explicit discussions, and consider the OPE in coordinate space where the Wilson coefficient is a function of $x_Q^2$. In $D=6-2\epsilon$ dimension, the mass dimension of a scalar field of cubic theory is $[\phi]=2$. When both the two operators are set to be fundamental fields, since the lowest order of operator basis is a single fundamental field, we get the OPE in leading order term as
\begin{equation}
{\phi}(x){\phi}(0)~=~ c_0(x_Q^2)~{\phi}(0)+\cdots~,
\end{equation}
where the $c_0(x_Q^2)$ is the Wilson coefficient of leading order term. Counting the mass dimension of both sides, we get mass dimension $[c_0]=2$, so $c_0(x_Q^2)\sim \frac{1}{x_Q^2}$. Similarly, if one operator is ${\phi}^Q$ and the other is ${\bar{\phi}}$, the lowest order of operator basis is ${\phi}^{Q-1}$, then
\begin{equation}
{\phi}^{Q}(x){\bar{\phi}}(0)~=~ c_0(x_Q^2)~{\phi}^{Q-1}(0)+\cdots~.\label{eqn:OPE-LO-6D-FF}
\end{equation}
The equality of mass dimension in both sides leads to $[c_0]=4$, so the leading order term is $c_0\sim \frac{1}{(x_Q^2)^2}$. The same discussion can be applied to the quartic theory in $D=4-2\epsilon$ dimension, where the mass dimension of scalar field is $[\phi]=1$. Again if both the two operators are fundamental fields, and now the lowest order of operator basis is ${\phi}^2$, then the OPE equation leads to
\begin{equation}
{\phi}(x){\phi}(0)~=~ c_0(x_Q^2)~{\phi}^2(0)+\cdots~~~~\to~~~ [c_0]=0~~~\to~~~c_0\sim 1~.
\end{equation}
If instead one operator is ${\phi}^Q$, the equality of mass dimension in both sides of eqn.(\ref{eqn:OPE-LO-6D-FF}) requires $[c_0]=2$, and we have $c_0\sim \frac{1}{x_Q^2}$. All the behaviors of leading order terms are in agreement with the previous discussions of large momentum expansion.

\subsection{Expansion of the correlation functions and the computation of $Z$-factors}
\label{subsec:full-expansion}

From the above general discussions we know that, the Wilson coefficient of OPE is proportional to the hard graphs in the large momentum expansion, and the latter are computed by two-point integrals. This idea is hold in general quantum field theories for various operators. As an illustration, let us apply this idea to the computation of $Z$-factors of the $\phi^Q$ operator in scalar theories\footnote{The determination of $Z$-factors relies on the UV finiteness of the Wilson coefficients in the OPE, as shown in eqn.(\ref{eqn:OPE-FF}). Since the OPE is valid for general quantum field theories, the validation of OPE approach for renormalization computations will not depend on the details of the operators. However, for general theories and operators, we might need IBP to evaluate the integrals.}. The anomalous dimension of the ${\phi}^Q$ operator can be extracted from the following correlation function,
\begin{equation}
\Bigl\langle {\phi}^Q_R(x_0)\phi(x_1)\cdots\phi(x_Q)\Bigr\rangle~,
\end{equation}
where the ${\phi}^Q_R$ is the renormalized operator. In order to compute this correlation function, we need to work out the corresponding Feynman rules. The renormalized operator is related to the normal operator ${\phi}^Q$ through the bare operator ${\phi}_0^Q$ as,
\begin{equation}
{\phi}^Q_R=Z_{\phi^Q}{\phi}_0^Q= \widetilde{Z}_{\phi^Q}{\phi}^Q~,
\end{equation}
where we have used ${\phi}_0^Q=Z_{\phi}^{Q/2}{\phi}^Q$ and define $\widetilde{Z}_{\phi^Q}:=Z_{\phi^Q}Z_{\phi}^{Q/2}$. Therefore, we can determine the Feynman rule of ${\phi}_R^Q$-$Q\phi$ vertex as $\widetilde{Z}_{\phi^Q}$. The anomalous dimension of the ${\phi}^Q$ will be determined by the $Z_{\phi^Q}$. Furthermore, each loop propagator can be combined with counterterms to form an {\sl effective} propagator as
\begin{equation}
\frac{i}{\ell^2}+\frac{i}{\ell^2}i(Z_{\phi}-1)\ell^2\frac{i}{\ell^2}
+\cdots
=\frac{iZ_{\phi}^{-1}}{\ell^2}~,
\end{equation}
which defines the Feynman rule for loop propagators. The Feynman rule for a scalar vertex reads $-iZ_gg\tilde{\mu}^{2\epsilon}=-ig_0Z_{\phi}^2$, where $\mu$ is the scale parameter. The dependence of $Z$-factors in the complete correlation function can be computed by counting all the $Z$-factors from the operator vertex, loop propagators and scalar vertices. For example, let us consider the quartic scalar model, and suppose a $L$-loop Feynman diagram has one ${\phi}^Q_R$-$Q\phi$ vertex,  $n_{v}$ $\phi^4$ vertices, and $n_e$ loop propagators. Then counting the number of vertices, edges and loops we get the relations
\begin{equation}
Q+4n_{v}=2n_{e}+Q~~~,~~~(n_v+1)-n_e+L=1~~~\to~~~n_v=L~~~,~~~n_e=2L~.
\end{equation}
So the diagram comes with the following combination of Z factors,
\begin{equation}
\widetilde{Z}_{\phi^Q}Z_{\phi}^{-n_e}(g_0Z_{\phi}^2)^{n_v}
=\widetilde{Z}_{\phi^Q}g_0^L~.
\end{equation}
For $L$-loop Feynman diagram, the $Z$-factor appears as an overall factor $\widetilde{Z}_{\phi^Q}$ universally, and the loop order only shows up in the couplings. Hence the all-loop correlation function can be written as
\begin{equation}
\Bigl\langle {\phi}^Q_R(x_0)\phi(x_1)\cdots\phi(x_Q)\Bigr\rangle=\widetilde{Z}_{\phi^Q}\sum_{L=0}^{\infty} g_0^L G_Q^{(L)}(x_0,x_1,\ldots, x_Q)
:= \widetilde{Z}_{\phi^Q}G_Q(g_0; x_0,x_1,\ldots, x_Q)~,\label{eqn:FF-full-expansion}
\end{equation}
where the $G_Q(g_0;x_0,\cdots, x_Q)$ is the all-loop bare correlation function. 

For general quantum field theory, the correlation function has the same structure as
\begin{equation}
\Bigl\langle \widehat{\mathcal{O}}_R(x_0)\phi(x_1)\cdots\phi(x_n)\Bigr\rangle
=\widetilde{Z}_{\mathcal{O}}G_{\mathcal{O}}(g_{i0};x_0,x_1,\ldots, x_n)~,\label{eqn:OPE-general-operator}
\end{equation}
where the $g_{i0}$ represents one or more bare coupling constants of the theory, and all the coupling constants $g_i$'s are replaced by the bare coupling constants $g_{i0}$'s at the end. In fact, the correlation functions of fundamental fields also have similar structures. For example, in $\phi^4$ theory, we have
\begin{equation}
\Bigl\langle\phi(x_1)\phi(x_2)\Bigr\rangle
=Z_{\phi}G(g_{0};x_1, x_2)~~~,~~~\Bigl\langle\phi(x_1)\phi(x_2)\phi(x_3)\phi(x_4)\Bigr\rangle
=Z_{\phi}^2G(g_{0};x_1, x_2,x_3,x_4)~,
\end{equation}
and the $Z$-factors appear as a universal overall factor. We will shortly see that, the UV finiteness of correlation functions determines the overall $Z$-factors, and also gives constraints to the lower-loop correlations of $g_{i0}$.

\begin{figure}
\centering
\begin{tikzpicture}
\node [] at (2,0) {
    \begin{tikzpicture}
    \draw [] (0,0)--(2,1.5) (0,0)--(2.5,1) (0,0)--(2.75,0) (0,0)--(2.5,-1);
    \draw [very thick] (0,0)--(2,-1.5);
    \draw [fill=black] (2.25, 0.7) circle [radius=0.05] (2.3, 0.5) circle [radius=0.05] (2.32,0.25) circle [radius=0.05] (2.25, -0.7) circle [radius=0.05] (2.3, -0.5) circle [radius=0.05] (2.32,-0.25) circle [radius=0.05];
    \draw [fill=lightgray,thick] (1.25,0) ellipse [x radius=0.75, y radius=1.75];
    \draw [fill=white] (0, 0) circle [radius=0.1];
    \draw [] (-0.07,-0.07)--(0.07,0.07) (-0.07,0.07)--(0.07,-0.07);
    \node [] at (1.25,0) {$L$};
    \node [above left] at (0,0) {$\phi^Q$};
    \node [right] at (2,1.5) {$p_1$};
    \node [right] at (2.5,1) {$p_2$};
    \node [right] at (2.5,-1) {$p_{Q-1}$};
    \node [right] at (2,-1.5) {$p_Q$};
    \end{tikzpicture}
};
\node [] at (8.5,0) {
    \begin{tikzpicture}
    \draw [] (0,0)--(1,0.75) (0,0)--(1.25,0.5) (0,0)--(1.5,0) (0,0)--(1.25,-0.5);
    \draw [dashed, thick] (1,0.75)--(2,1.5) (1.25,0.5)--(2.5,1) (1.5,0)--(2.75,0) (1.25,-0.5)--(2.5,-1);
    \draw [very thick] (0,0)--(2,-1.5);
    \draw [fill=lightgray,thick] (1.25,0) ellipse [x radius=0.75, y radius=1.75];
    \draw [fill=white] (0, 0) circle [radius=0.1];
    \draw [] (-0.07,-0.07)--(0.07,0.07) (-0.07,0.07)--(0.07,-0.07);
    \node [] at (1.25,0) {$L_1$};
    \node [above left] at (0,0) {$\phi^Q$};
    \node [right] at (2,-1.5) {$p_Q$};
    \end{tikzpicture}
};
\node [] at (13.5,0) {
    \begin{tikzpicture}
    \draw [] (0,0)--(2,1.5) (0,0)--(2.5,1) (0,0)--(2.75,0) (0,0)--(2.5,-1);
    \draw [] (0,0)--(1,-0.75);
    \draw [fill=black] (2.25, 0.7) circle [radius=0.05] (2.3, 0.5) circle [radius=0.05] (2.32,0.25) circle [radius=0.05] (2.25, -0.7) circle [radius=0.05] (2.3, -0.5) circle [radius=0.05] (2.32,-0.25) circle [radius=0.05];
    \draw [fill=lightgray,thick] (1.25,0) ellipse [x radius=0.75, y radius=1.75];
    \draw [fill=white] (0, 0) circle [radius=0.1];
    \draw [] (-0.07,-0.07)--(0.07,0.07) (-0.07,0.07)--(0.07,-0.07);
    \node [] at (1.25,0) {$L_2$};
    \node [above left] at (0,0) {$\phi^{Q-1}$};
    \node [right] at (2,1.5) {$p_1$};
    \node [right] at (2.5,1) {$p_2$};
    \node [right] at (2.5,-1) {$p_{Q-1}$};
    \end{tikzpicture}
};
\node [] at (5, 0) {$\to$};
\node [] at (6,0) {$\sum$};
\node [] at (10.75,0) {$\circ$};
\end{tikzpicture}
\caption{The pictorial representation of the large momentum expansion for the ${\phi^Q}\to Q\phi$ form factor, where the momentum of the operator and an external momentum $p_Q$ are taken to the large limit. By applying the minimal cuts to the graph, we extract the leading order contribution of expansion. The expansion terms are described by the hard graphs of two-point functions with large momenta, and the soft graphs of lower-point correlation functions with on-shell soft momenta.}\label{fig:phiQ-expansion}
\end{figure}
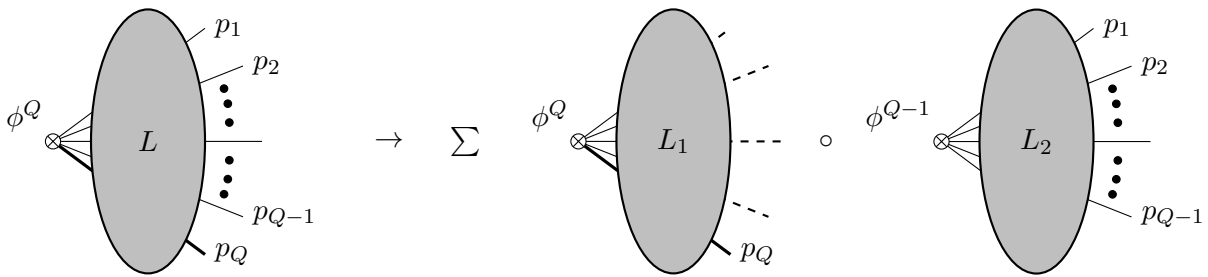

Now let us apply the large momentum expansion to the correlation function eqn.(\ref{eqn:FF-full-expansion}) and extract the leading order contribution of the asymptotic expansion. This can be realized by applying the minimal cuts to the Feynman graphs. When taking the momentum of the operator and an external momenta $p_Q$ to the large limit, and assuming $p_Q\gg p_i$, $\forall i\neq Q$, we can asymptotically expand the correlation function around the large $p_Q$, as sketched in Fig.(\ref{fig:phiQ-expansion}). For the ${\phi}^Q$ operator, we should consider all possible minimal $(Q-1)$-cuttings of the original Feynman graph. Consequently, the expansion terms are given by products of hard graphs of two-point functions and soft graphs of ${\phi^{Q-1}}\to (Q-1)\phi$ form factors with on-shell soft momenta. In coordinate space, assuming the coordinate of operator is $x_0$ and that of large momentum $p_Q$ is $x_Q$, the large momentum expansion is realized by taking the $x_Q\to x_0$ limit.

Applying the large momentum expansion on the RHS of eqn.(\ref{eqn:FF-full-expansion}), the $L$-loop correlation function is expanded as
\begin{equation}
G_Q^{(L)}(x_0,x_1,x_2,\ldots,x_Q)\cong\sum_{i=0}^{L}G_{Q}^{(L-i)}(x_0,\del{x}_1,\ldots,\del{x}_{Q-1},x_Q)
G_{Q-1}^{(i)}(x_0,x_1,x_2,\ldots, x_{Q-1})~,\label{eqn:FF-L-expansion}
\end{equation}
where the $\del{x}_i$ represents the removal of a corresponding external leg. The summation is over all possible minimal $(Q-1)$-cuttings that split Feynman graph to the 1PI sub-graphs, producing the leading order terms in the large momentum expansion. Substituting above expansion back to the RHS of eqn.(\ref{eqn:FF-full-expansion}) and switching the order of summations, we get
\begin{eqnarray}
&&\widetilde{Z}_{\phi^Q}\sum_{i=0}^{\infty}\sum_{L-i=0}^{\infty}g_0^L G_{Q}^{(L-i)}(x_0,\del{x}_1,\ldots,\del{x}_{Q-1},x_Q)
{G}_{Q-1}^{(i)}(x_0,x_1,x_2,\ldots, x_{Q-1})\\
&&~~~~~~~~~~~~~~~=\widetilde{Z}_{\phi^Q}\left(\sum_{i=0}^{\infty}g_0^i{G}_{Q-1}^{(i)}(x_0,x_1,x_2,\ldots, x_{Q-1})\right)\left(\sum_{L-i=0}^{\infty} g_0^{L-i}{G}_{Q}^{(L-i)}(x_0,\del{x}_1,\ldots,\del{x}_{Q-1},x_Q)\right)~.\nonumber
\end{eqnarray}
Expression in the first bracket times $\widetilde{Z}_{\phi^{Q-1}}$ is the full correlation function of the ${\phi}^{Q-1}\to (Q-1)\phi$. Hence we get the relation
\begin{equation}
\Bigl\langle {\phi}^Q_R(x_0)\phi(x_1)\cdots\phi(x_Q)\Bigr\rangle\cong\frac{\widetilde{Z}_{\phi^Q}}{\widetilde{Z}_{\phi^{Q-1}}}\Bigl\langle {\phi}^{Q-1}_R(x_0)\phi(x_1)\cdots\phi(x_{Q-1})\Bigr\rangle\left(\sum_{j=0}^{\infty} g_0^{j}{G}_{Q}^{(j)}(x_0,\del{x}_1,\ldots,\del{x}_{Q-1},x_Q)\right)~.\label{eqn:LME-LO}
\end{equation}
Since both correlation functions are UV finite, the expansion coefficient should also be UV finite in the dimensional regularization parameter $\epsilon$. The finiteness condition constraints the ratio of $Z$-factors by relation as,
\begin{equation}
\frac{\widetilde{Z}_{\phi^Q}}{\widetilde{Z}_{\phi^{Q-1}}}\left(\sum_{j=0}^{\infty} g_0^{j}{G}_{Q}^{(j)}(x_0,\del{x}_1,\ldots,\del{x}_{Q-1},x_Q)\right):=
\frac{\widetilde{Z}_{\phi^Q}}{\widetilde{Z}_{\phi^{Q-1}}}{G}_{Q}(g_0;x_0,\del{x}_1,\ldots,\del{x}_{Q-1},x_Q)={\rm UV~finite}~.\label{eqn:ope-ratio}
\end{equation}
From the UV finiteness condition and the $\epsilon$ expansion results of two-point functions, the ratio of $Z$-factors can be completed determined. The minimal cuts of the large momentum expansion extract the leading order terms of the $x_Q^2$ expansion, while there is also $\ln^k(x_Q^2)$-dependence in the result. Since
each rank of logarithms is also independent, we can select the coefficients of leading order logarithm $\ln^0(x_Q^2)$ in the result of leading $x_Q^2$ contribution to fit the ratio of $Z$-factors. The coefficients of higher rank logarithms also generate system of algebraic equations from UV finiteness condition, and they can solve lower-loop beta functions. The above relation is valid for any loop order and $Q$. Once the two-point functions are computed, the $Z$-factors can be determined recursively by UV finiteness conditions loop by loop.

Comparing eqn.(\ref{eqn:LME-LO}) with the OPE expansion eqn.(\ref{eqn:OPE-FF}), we see that the eqn.(\ref{eqn:ope-ratio}) is exactly the Wilson coefficient. In the OPE, the Wilson coefficient is singular in $x_Q^2$ but finite in dimensional regularization parameter $\epsilon$. This is the origin of UV finiteness condition. It is now clear that the leading order term of $x_Q^2$ series in the large momentum expansion corresponds to the Wilson coefficient of the leading order operator basis in OPE. We want to emphasize again that, in the above framework of determining $Z$-factors, the only input is the two-point propagator-type integrals of hard field and the ${\phi^Q}$ operator, and the corresponding Feynman integral is an one scale function. The Feynman graphs of two-point integrals can be generated by removing $(Q-1)$ external legs $x_1,x_2,\ldots,x_{Q-1}$ from the original Feynman graphs of the ${\phi^Q}\to Q\phi$ correlation functions. Similarly, when computing the anomalous dimension of field, mass or couplings, we need to deal with three-point and four-point correlation functions. In the framework eqn.(\ref{eqn:ope-ratio}), what we really need to compute is the two-point integrals generated by graphs of three or four-point functions with the removal of one or two external legs. For the scalar theories, we choose to compute these two-point integrals by graphical function method, which can be evaluated by running \verb!TwoPoint! command in \verb!HyperlogProcedures! package \cite{HP} with the coordinate of the ${\phi}^Q$ operator locating at point 0 and the coordinate of hard field locating at point 1.

In the dimensional regularization, the $L$-loop two-point function ${G}_{Q}^{(L)}(x_0,\del{x}_1,\ldots,\del{x}_{Q-1},x_{Q})$ can be computed as an expansion in the dimensional regulator $\epsilon$ with the leading order pole $1/\epsilon^L$. In order to correctly generate equations of UV finiteness condition, the lower-loop two-point functions should be computed to appropriate orders of rational $\epsilon$ terms. Generally in the \verb!HyperlogProcedures! package, computing integrals to one order higher $\epsilon$ terms would cause several times greater in the time consumption. This is a limit for pushing the computation of renormalization functions towards the next-loop challenge. The determination of the $\widetilde{Z}_{\phi^Q}$ can be done recursively with the increasing parameter $Q$ and the loop order $L$. In fact, the finiteness conditions allow us to completely fix all ratios of $Z$-factors with different values of $Q$ and $L$ simultaneously. Hence each $\widetilde{Z}_{\phi^Q}$ is determined when all ratios are solved. Then we can follow the standard textbook method to compute the anomalous dimensions, {\sl etc}.

The above method of computing $Z$-factors applies to the correlation functions of fundamental fields with the ${\phi^Q}$ operator inserted in either 6-dimensional cubic scalar theory or 4-dimensional quartic scalar theory. It also applies to the correlation functions of fundamental fields. For the 6-dimensional cubic scalar theory, the minimal number of cuts is 1. This allows us to compute $Z$-factors from the related three-point functions by removing one external leg. For the 4-dimensional quartic scalar theory, the minimal number of cuts is 2, and it allows us to compute $Z$-factors from the corresponding four-point functions with two external legs removed.

\section{The anomalous dimension of the $\phi^Q$ operator in $O(N)$ cubic scalar theory}
\label{sec:RG-example}

As a demonstration of the OPE method for computing renormalization functions, in this section we will compute the anomalous dimensions of ${\phi}^Q$ operators in cubic scalar theory to $4$-loop, and verify the validation and effectiveness of our algorithm.

\subsection{The $O(N)$ model and the ${\phi}^Q$ operator}

The renormalization of scalar $\phi^4$ theory with $O(N)$ symmetry has long been the focus of study, because the  critical exponents in different statistical systems can be obtained from the scaling dimensions of operators at the Wilson-Fisher fixed point \cite{Wilson:1983xri,PhysRevB.4.3184,Wilson:1971dc}. The beta function and anomalous dimensions of some operators, {\sl e.g.}, $\phi$ and $\phi^Q$, are computed to 6-loop using the $R^*$ operation method \cite{Kompaniets:2017yct}, and recently to 7-loop using the graphical functions \cite{Schnetz:2022nsc}. When some parameters are taken to the large limit, physical quantities can be expanded around the limit by Taylor series, and usually the expansion results are comparable to the results computed by other approaches. The large $N$ expansion of the scaling dimensions of various operators to the $\mathcal{O}(\frac{1}{N^3})$ order can be obtained by the conformal bootstrap method in an entire range of spacetime dimensions \cite{Vasilev:2004yr}.
Meanwhile, the semiclassical method
has been generalized to compute the scaling dimension of the $\phi^Q$-type operators in the large charge limit for various theories to all loops, including theories with scalars and gauge fields \cite{Badel:2019oxl,Antipin:2020abu,Antipin:2020rdw,Jack:2021ypd,Antipin:2021akb,Jack:2021lja,Antipin:2022naw,Antipin:2022hfe}. But to obtain the complete results of anomalous dimensions, we still need the help of the perturbation method.

In \cite{Fei:2014yja}, it is proposed that the $O(N)$ quartic scalar theory at its UV fixed point is equivalent to the following $O(N)$ cubic scalar theory at its IR fixed point,
\begin{equation}
\mathcal{L}=\frac{1}{2}(\partial\phi_i)^2+\frac{1}{2}(\partial\sigma)^2-\frac{g}{2}\sigma\phi_i^2-\frac{h}{6}\sigma^3~,
\end{equation}
with two kinds of scalar fields $\sigma$ and $\phi_i,i=1,\ldots,N$.  The equivalence is further verified to 4 and 5-loops \cite{Fei:2014xta,Gracey:2015tta,Kompaniets:2021hwg,Borinsky:2021jdb} for the $\phi$ and $\phi^2$ operators in large $N$ limit, and it is verified to 3-loop for the anomalous dimensions of the $\phi^Q$ operators \cite{Jack:2021ziq}. In this section, we will compute the anomalous dimensions of the $\phi^Q$ operators in six-dimensional cubic theory to 4-loops.

The operator $\phi^Q$ is defined as the symmetric traceless operator
\begin{equation}
\phi^Q:= T_{i_1\cdots i_Q}\phi^{i_1}\cdots\phi^{i_Q}~.
\end{equation}
As discussed in \cite{Jin:2022nqq}, it is convenient to choose
\begin{equation}
T_{i_1\cdots i_Q}=\frac{1}{Q!}q_{i_1}\cdots q_{i_Q}~,
\end{equation}
so that the operator becomes
\begin{equation}
\phi^Q:= \frac{1}{Q!}\varphi^Q~,
\end{equation}
in which $q$ is an auxiliary $O(N)$ zero-norm complex vector satisfying $q^2=0$, $q\cdot \bar{q}=1$, and the $\varphi$ is a redefinition of scalar field $\varphi:=q\cdot \phi$. With this definition, we can strip off $O(N)$ indices from the correlation function, and compute a simpler $\varphi^Q\to Q\varphi$ correlation function.

As proven in \S\ref{subsec:full-expansion}, the ratio $\widetilde{Z}_{\varphi^Q}/\widetilde{Z}_{\varphi^{Q-1}}$ can be determined from the UV finiteness of the Wilson coefficient $\mathcal{C}_Q$,
\begin{equation}
\mathcal{C}_Q(x_Q)=\frac{\widetilde{Z}_{\varphi^Q}}{\widetilde{Z}_{\varphi^{Q-1}}}\left(\sum_{j=0}^{L} g_0^{j}{G}_{Q}^{(j)}(x_0,\del{x}_1,\ldots,\del{x}_{Q-1},x_Q)\right)={\rm UV~finite}~,\label{eqn:def-ratio}
\end{equation}
up to the desired loop order $L$. The above relation requires us to compute the two-point function ${G}_{Q}$, which is generated from the $\varphi^Q\to Q\varphi$ correlation functions by removing $(Q-1)$ external legs, leaving only the off-shell operator and one external leg as large momenta in the large momentum expansion.

\subsection{Treatments on the Feynman graphs}

When the charge $Q$ and the loop order $L$ are large, it can be difficult to generate all contributing Feynman graphs. This has already been a problem for 4-loop case when $Q$ takes the value of 4 or 5. Hence it is better to simplify the problem of generating Feynman graphs as much as possible by using properties of Lagrangian and graphs. Here are some techniques that we adopted to reformulate Feynman graphs before computing integrals.

\subsection*{Summing over internal states}

One difficulty of the considered cubic theory is that we need to consider two different scalar fields, which increases the number of Feynman graphs drastically. We circumvent this difficulty by combining two scalar fields into a single $(N+1)$ component scalar field $\Phi$ as,
\begin{equation}
\Phi_{I}=(\Phi_0, \Phi_i,\ldots,\Phi_N):=(\sigma, \varphi_1,\ldots,\varphi_N)~.
\end{equation}
Then the cubic potential terms can be written as
\begin{equation}
V(\Phi)=\frac{1}{6}\lambda_{IJK}\Phi_I\Phi_J\Phi_K~.
\end{equation}
The coupling $\lambda_{IJK}$ is symmetric, and the original couplings $h, g$ have been defined as the components of $\lambda_{IJK}$,
\begin{equation}
\lambda_{000}=h~~~,~~~\lambda_{0ij}=g\delta_{ij}~~,~~\forall i,j\neq 0~,~~~
\end{equation}
and all other components zero. In this definition, the contributions of different $\sigma$, $\varphi_i$'s internal state configurations of the same topology can be combined into a {\sl coupling factor} $\alpha(g,h,N)$, which can be obtained from the products of $\lambda_{IJK}$'s. For example, let us consider an 1-loop 2-point Feynman graph of the $\Phi$ field as shown below,
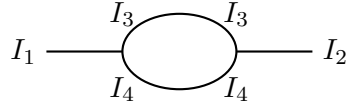
\begin{center}
\begin{tikzpicture}
  \draw [thick] (0,0)--(1,0) (2.5,0)--(3.5,0);
  \draw [thick] (1.75,0) ellipse [x radius=0.75, y radius=0.5];
  \node [left] at (0,0) {$I_1$};
  \node [right] at (3.5,0) {$I_2$};
  \node [ ] at (1,0.5) {$I_3$};
  \node [] at (1,-0.5) {$I_4$};
  \node [] at (2.5,0.5) {$I_3$};
  \node [] at (2.5,-0.5) {$I_4$};
\end{tikzpicture}
\end{center}
The coupling factor for this graph is a product of two cubic vertices, which reads
\begin{equation}
\alpha_{I_1I_2}(g,h,N)=\sum_{I_3,I_4=0}^{N}\lambda_{I_1I_3I_4}\lambda_{I_2I_4I_3}~,
\end{equation}
where $I_1$ and $I_2$ are indices of external legs. If the external legs are $\sigma$ fields, we set $I_1=I_2=0$ and obtain
\begin{equation}
\alpha_{00}(g,h,N)=\lambda_{000}^2+\sum_{i=1}^{N}\lambda_{0ii}^2=h^2+Ng^2~.
\end{equation}
In the graphs of the $\sigma$ and $\varphi_i$ fields, this coupling factor just shows the fact that, the graph with $\sigma$ fields as internal propagators and those with $\varphi$'s fields as internal propagators can be combined together, with a scalar integral of the topology as overall factor. While if the external legs are $\varphi$ fields, the only possible internal state configuration is a $\sigma$ propagator and a $\varphi_i$ propagator. By setting $I_1=i_1,I_2=i_2$ and computing the coupling factor, we get
\begin{equation}
\alpha_{i_1i_2}(g,h,N)=\sum_{I_4=1}^{N}\lambda_{i_10I_4}\lambda_{i_2I_40}
+\sum_{I_3=1}^{N}\lambda_{i_1I_30}\lambda_{i_20I_3}=2g^2\delta_{i_1i_2}~,
\end{equation}
which is as expected from the graphs of $\sigma,\varphi_i$ fields. The coupling factor of $\varphi^Q$ correlation functions can be evaluated in the same manner. In this treatment, the couplings and the $N$-dependence of graphs with different internal states but the same topology can be organized as a single coupling factor $\alpha(g,h,N)$. We only need to compute the integrals of independent topologies, getting rid of the abundant graphs by assigning internal states.

\subsection*{Reducible and irreducible graphs}

\begin{figure}
\centering
\begin{tikzpicture}
\node [] at (2,0) {
    \begin{tikzpicture}
    \draw [thick] (0,0)--(2.5,-1) (0,0)--(2.5,1);
    \draw [fill=black] (2.19, -0.24) circle [radius=0.03] (2.19, 0.24) circle [radius=0.03];
    \draw [fill=lightgray] (1.5,0) ellipse [x radius=0.35, y radius=0.75];
    \draw [fill=white] (0, 0) circle [radius=0.1];
    \draw [] (-0.07,-0.07)--(0.07,0.07) (-0.07,0.07)--(0.07,-0.07);
    \end{tikzpicture}
};
\node [] at (10,0) {
    \begin{tikzpicture}
    \draw [thick] (0,0)--(0,-2.5);
    \draw [thick] (0,0)--(2,-2) (0,0)--(2.5,-0.83) (0,0)--(2.5,0.83) (0,0)--(2,2) (0,0)--(0,2.5);
    \draw [fill=black] (2.19, -0.24) circle [radius=0.03] (2.19, 0.24) circle [radius=0.03] (0.89, -0.46) circle [radius=0.03] (0.81, -0.59) circle [radius=0.03] (0.52, -1.93) circle [radius=0.03] (1, -1.73) circle [radius=0.03]  (0.89, 0.46) circle [radius=0.03] (0.81, 0.59) circle [radius=0.03] (0.52, 1.93) circle [radius=0.03] (1, 1.73) circle [radius=0.03];
    \draw [fill=lightgray] (1.5,0) ellipse [x radius=0.35, y radius=0.75];
    \draw [fill=lightgray,rotate=-67] (1.5,0) ellipse [x radius=0.35, y radius=0.85];
    \draw [fill=lightgray,rotate=67] (1.5,0) ellipse [x radius=0.35, y radius=0.85];
    \draw [fill=white] (0, 0) circle [radius=0.1];
    \draw [] (-0.07,-0.07)--(0.07,0.07) (-0.07,0.07)--(0.07,-0.07);
    \end{tikzpicture}
};
\node [] at (10, -3) {(b) reducible graph};
\node [] at (2,-3) {(a) irreducible graph};
\end{tikzpicture}
\caption{(a) For the irreducible graphs, all external legs are connected to the operator vertex via the same loop sub-graph. (b) For the  reducible graphs, external legs are distributed in two or more different loop sub-graphs connecting to the operator vertex. The reducible graph can be expressed as the products of several irreducible graphs.}\label{fig:irreducible-graph}
\end{figure}
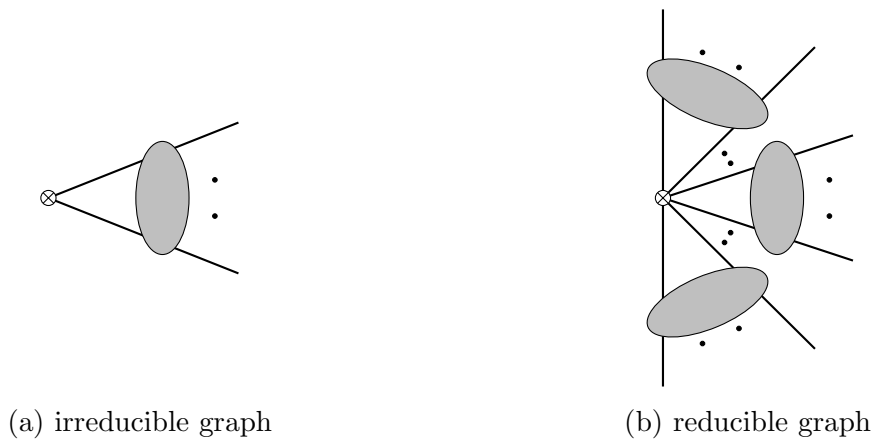

The Feynman graphs of two-point function ${G}_{Q}(x_0,\del{x}_1,\ldots,\del{x}_{Q-1},x_Q)$ are generated from Feynman graphs of the $\varphi^Q\to Q\varphi$ correlation function by removing $(Q-1)$ external legs. We will show that, after the removal of external legs we only need to compute a part of the original graphs. All the Feynman graphs of the $\varphi^Q\to Q\varphi$ correlation function can be classified as irreducible graphs and reducible graphs, according to the way of external legs connecting to the operator vertex. An external leg can attach to the operator vertex directly or via a loop sub-graph. The directly connected leg gives an identity in the integrand, which could be trivially omitted. We can now classify the graphs with external legs connected to operator vertex via loop sub-graphs. If all external legs are connected to the same loop sub-graph, we define it as an {\sl irreducible graph}. While if all external legs are distributed in more than one loop sub-graphs, we call it a {\sl reducible graph}, as illustrated in Fig.(\ref{fig:irreducible-graph}).

From the graphs we can infer that, the integrand of the reducible graph can be decomposed into the integrands of several irreducible graphs. The $L$-loop decomposition formula for arbitrary $Q$ is given in \cite{Jin:2022nqq}, and here we present an all-loop expression,
\begin{equation}\label{eqn:form-factor-exp}
G_Q(x_0,x_1,\ldots,x_Q)= \sum_{k=0}^{[Q/2]}\sum_{{2\leq n_1\leq \cdots \leq n_k \atop n_1+\cdots+n_k\leq Q}}\frac{Q!}{(Q-\sum_{i=1}^kn_i)!}
\frac{1}{S_{n_1\cdots n_k}}\prod_{i=1}^k\frac{1}{n_i!}\mathtt{F}_{n_i}~,
\end{equation}
where $k$ is a label denoting the number of loop sub-graphs, and $k=0$ stands for contributions of tree-level graphs. The $S_{n_1\cdots n_k}$ is a symmetry factor, with the definition that when there are duplicated $n_i$'s it gets a non-trivial factorial value for each duplication. For instance $S_{11444}=2!3!$ with $2!$ from double duplicated indices $1$ and $3!$ from triple duplicated indices $4$. The $\mathtt{F}_{n_i}$ denotes the integrand of the irreducible graph with $n_i$ external legs\footnote{The graphs of correlation function could have external legs connected directly to the operator vertex besides the irreducible graph. Since graphs which are related by permutations of external legs produce the same local UV counterterm, we can formally write the permutation related graphs as copies of one graph. For instance, the following four graphs
\begin{center}
\begin{tikzpicture}
\node [] at (1,0) {
    \begin{tikzpicture}
    \draw [thick] (0,0)--(1.12,0) (0,0)--(1,0.5) (0,0)--(1,-0.5) (0,0)--(0,0.5);
    \draw [fill=lightgray] (0.5,0) ellipse [x radius=0.15, y radius=0.5];
    \draw [fill=white] (0, 0) circle [radius=0.1];
    \draw [] (-0.07,-0.07)--(0.07,0.07) (-0.07,0.07)--(0.07,-0.07);
    \node [above] at (0,0.5) {$x_1$};
    \node [right] at (1,0.5) {$x_2$};
    \node [right] at (1.12,0) {$x_3$};
    \node [right] at (1,-0.5) {$x_4$};
    \end{tikzpicture}
};
\node [] at (4,0) {
    \begin{tikzpicture}
    \draw [thick] (0,0)--(1.12,0) (0,0)--(1,0.5) (0,0)--(1,-0.5) (0,0)--(0,0.5);
    \draw [fill=lightgray] (0.5,0) ellipse [x radius=0.15, y radius=0.5];
    \draw [fill=white] (0, 0) circle [radius=0.1];
    \draw [] (-0.07,-0.07)--(0.07,0.07) (-0.07,0.07)--(0.07,-0.07);
    \node [above] at (0,0.5) {$x_2$};
    \node [right] at (1,0.5) {$x_1$};
    \node [right] at (1.12,0) {$x_3$};
    \node [right] at (1,-0.5) {$x_4$};
    \end{tikzpicture}
};
\node [] at (7,0) {
    \begin{tikzpicture}
    \draw [thick] (0,0)--(1.12,0) (0,0)--(1,0.5) (0,0)--(1,-0.5) (0,0)--(0,0.5);
    \draw [fill=lightgray] (0.5,0) ellipse [x radius=0.15, y radius=0.5];
    \draw [fill=white] (0, 0) circle [radius=0.1];
    \draw [] (-0.07,-0.07)--(0.07,0.07) (-0.07,0.07)--(0.07,-0.07);
    \node [above] at (0,0.5) {$x_3$};
    \node [right] at (1,0.5) {$x_1$};
    \node [right] at (1.12,0) {$x_2$};
    \node [right] at (1,-0.5) {$x_4$};
    \end{tikzpicture}
};
\node [] at (10,0) {
    \begin{tikzpicture}
    \draw [thick] (0,0)--(1.12,0) (0,0)--(1,0.5) (0,0)--(1,-0.5) (0,0)--(0,0.5);
    \draw [fill=lightgray] (0.5,0) ellipse [x radius=0.15, y radius=0.5];
    \draw [fill=white] (0, 0) circle [radius=0.1];
    \draw [] (-0.07,-0.07)--(0.07,0.07) (-0.07,0.07)--(0.07,-0.07);
    \node [above] at (0,0.5) {$x_4$};
    \node [right] at (1,0.5) {$x_1$};
    \node [right] at (1.12,0) {$x_2$};
    \node [right] at (1,-0.5) {$x_3$};
    \end{tikzpicture}
};
\end{tikzpicture}
\end{center}
can be formally written as
\begin{equation}
4\mathtt{F}_{3}:=\frac{1}{x_1^{2\lambda}}\mathtt{F}_{3}(x_2,x_3,x_4)
+\frac{1}{x_2^{2\lambda}}\mathtt{F}_{3}(x_1,x_3,x_4)
+\frac{1}{x_3^{2\lambda}}\mathtt{F}_{3}(x_1,x_2,x_4)
\frac{1}{x_4^{2\lambda}}\mathtt{F}_{3}(x_1,x_2,x_3)~,
\end{equation}
where $1/x_i^{2\lambda}$ is the Feynman rule for external leg in coordinate space.
}.

A reducible graph does not contribute to the Wilson coefficient $\mathcal{C}_Q(x_Q)$ defined in eqn.(\ref{eqn:def-ratio}). Because when external legs $x_1,\ldots,x_{Q-1}$ are removed, as can be inferred from Fig.(\ref{fig:irreducible-graph}), there would be at most one sub-graph with external leg $x_Q$, while other sub-graphs will inevitably become a scaleless snail graphs, and thus vanish. For instance, a reducible graph with four external legs distributing in two loop sub-graphs after removing three legs becomes
\begin{equation}
\mathtt{F}_{2}(x_1,x_2)\mathtt{F}_{2}(x_3,x_4)
~~~\rightarrow~~~ \mathtt{F}_{2}(\del{x}_1,\del{x}_2)\mathtt{F}_{2}(\del{x}_3,x_4)=0~,
\end{equation}
since $\mathtt{F}_{2}(\del{x}_1,\del{x}_2)$ is a scaleless integral.

An irreducible diagram can contribute to the $\mathcal{C}_Q(x_Q)$ only if $x_Q$ is not attached directly to the operator vertex. Otherwise after removing all the other legs, the loop sub-graph will be connected to the external momenta only via the operator vertex, and the whole graph becomes a scaleless snail diagram, as
\begin{equation}
\frac{1}{(x_{n+1}^2\cdots x_{Q}^2)^{\lambda}}\mathtt{F}_{n}(x_1,\cdots, x_n)
~~~\rightarrow~~~ \frac{1}{x_{Q}^{2\lambda}}\mathtt{F}_{n}(\del{x}_1,\cdots, \del{x}_n)=0~.
\end{equation}

Among all the terms in the decomposition eqn.(\ref{eqn:form-factor-exp}), the irreducible graphs are represented by terms with $k=1$, which is
\begin{equation}
\sum_{n=2}^{Q}\frac{Q!}{(Q-n)!}\frac{1}{n!}\mathtt{F}_n=\sum_{n=2}^{Q}C_{Q}^{n}\mathtt{F}_n~,
\end{equation}
where the $C_{Q}^{n}$ is the binomial coefficient. The summation is over all possible distributions of $Q$ external legs to the loop sub-graph and the operator vertex. For a given $\mathtt{F}_n$, the factor $C_{Q}^{n}$ just stands for the fact that because of the permutation symmetry, there are $C_Q^n$ copies of $\mathtt{F}_n$ contributing to the same local UV counterterm. However we should exclude graphs with $x_Q$ directly connected to the operator vertex, which has $C_{Q-1}^n$ copies. Then the remaining $C_{Q}^n-C_{Q-1}^n=C_{Q-1}^{n-1}$ copies of $\mathtt{F}_n$ have non-vanishing contributions to the Wilson coefficients. For example,
\begin{equation}
\frac{1}{(x_1^2\cdots x_{n-Q}^2)^{\lambda}}\mathtt{F}_{n}(x_{n-Q+1},\cdots, x_Q)
~~~\rightarrow~~~\mathtt{F}_{n}(\del{x}_{n-Q+1},\cdots, \del{x}_{Q-1}, x_Q)\neq 0~,
\end{equation}
where the resulting $\mathtt{F}_n$ is a two-point function with external leg $x_Q$ and the operator vertex. Eventually we obtain the contributions for two-point functions as
\begin{equation}\label{ratio-factor-exp}
 G_{Q}(x_0,\del{x}_{\sigma_1},\ldots,\del{x}_{\sigma_{Q-1}},x_Q)=\frac{1}{x_Q^{2\lambda}}+\sum_{n=2}^{Q}C_{Q-1}^{n-1}\mathtt{F}_{n}(\del{x}_{\sigma_1},\cdots, \del{x}_{\sigma_{n-1}}, x_Q)~,
\end{equation}
where the first term comes from the contributions of tree-level graphs, and the $x_{\sigma_1},\ldots,x_{\sigma_{n-1}}$ of $\mathtt{F}_n$ in the second term is an arbitrary subset of legs $x_1,\ldots, x_{Q-1}$, which should be removed. Now the original two-point function can be computed from all irreducible two-point graphs, which further reduces the number of graphs we need to compute.

In order to compute $\mathtt{F}_{n}(\del{x}_{\sigma_1},\cdots, \del{x}_{\sigma_{n-1}}, x_Q)$, firstly we generate all 1PI irreducible topologies contributing to the $\varphi^n\to n\varphi$ correlation functions $\mathtt{F}_{n}({x}_{\sigma_1},\cdots, {x}_{\sigma_{n-1}}, x_Q)$ with $n$ external legs and an operator vertex located at $x_0$, which will be denoted by $T_i$'s. Then for each $T_i$, we generate $n$ graphs by treating one external leg as special and compute the non-isomorphic graphs from them. The resulting graphs will be denoted by $\{T_{i,1},\ldots, T_{i,{m_i}}\}$. In coordinate space, each $T_{i,a}$ will be computed by removing all external legs but keeping only the special one, leaving a two-point graph with the operator vertex and the surviving external leg, and the two-point graph can be evaluated by graphical function method. Then the two-point function of irreducible graphs can be expressed as,
\begin{equation}
\mathtt{F}_{n}(\del{x}_{\sigma_1},\cdots, \del{x}_{\sigma_{n-1}}, x_Q)=(n-1)!\sum_{i=1}^{\#.{{\tiny topo.}}}\alpha_i(g,h,N)\sum_{a=1}^{m_i}
\frac{T_{i,a}}{S_{i,a}}~,\label{eqn:two-point-sum}
\end{equation}
in which the summation of $i$ runs over all 1PI irreducible topologies of $\mathtt{F}_n$, and $\alpha_i(g,h,N)$ is the coupling factor of topology $T_i$. The $S_{i,a}$ is the full symmetry factor of graph $T_{i,a}$, including the symmetry under permutation of vertices and internal propagators, as well as the symmetry under permutation of $\varphi_1,\ldots, \varphi_{n-1}$ external legs.

The number of two-point graphs for the $L$-loop $\varphi^Q\to Q\varphi$ correlation function is shown in Table.(\ref{table:phiQ-graph}), and these graphs will be computed by graphical function method in dimensional regularization scheme to get the $\epsilon$ series results. A sample computation can be found in Appendix \S\ref{appendix:sample_GF}.
\begin{table}
  \centering
  \begin{tabular}{|c|c|c|c|c|c|}
     \hline
      & $Q=2$ & $Q=3$ & $Q=4$ & $Q=5$ & Total\\ \hline
     $L=1$ & 1 & - & - & - & 1 \\ \hline
     $L=2$ & 7 & 4 & - & -  & 11 \\ \hline
     $L=3$ & 56 & 67 & 19 & - & 142 \\ \hline
     $L=4$ & 540 & 999 & 586 & 107 & 2232 \\
     \hline
   \end{tabular}
  \caption{The number of two-point graphs contributing to the two-point function of irreducible $L$-loop $\varphi^Q\to Q\varphi$ correlation function up to 4-loop. This is the number of graphs we need to compute by graphical function method.}\label{table:phiQ-graph}
\end{table}
%

\subsection{The anomalous dimensions}

In order to obtain the 4-loop correction of the $\gamma_{\varphi^Q}$, we computed the ratio ${\widetilde{Z}_{\varphi^Q}}/{\widetilde{Z}_{\varphi^{Q-1}}}$ up to 4-loop for $Q=2,3,4,5$. Using $\widetilde{Z}_{\varphi^1}=1$, we obtain $\widetilde{Z}_{\varphi^Q}$ for $Q=2,3,4,5$, which are suffice to fix $\widetilde{Z}_{\varphi^Q}$ to 4-loop for arbitrary $Q$. The anomalous dimension of $\varphi^Q$ is given by
\begin{equation}
\gamma_{\varphi^Q}=\frac{\partial \ln Z_{\varphi^Q}}{\partial \ln \mu}
=\frac{\partial \ln\Bigl(\widetilde{Z}_{\varphi^Q}Z_{\varphi}^{-\frac{Q}{2}}\Bigr)}{\partial \ln \mu}~.
\end{equation}
We computed $Z_{\varphi}$ and $Z_{\sigma}$ to 5-loop by evaluating $\langle\varphi\varphi\rangle$ and $\langle\sigma\sigma\rangle$ correlation functions using the two-point functions directly. The $Z_g$ and $Z_h$ to 4-loop are obtained as side-products, from the UV finiteness of $\ln (x^2_Q)$ terms in the two-point functions. Our results of $\gamma_{\phi}$, $\gamma_{\sigma}$, $\beta(g)$, $\beta(h)$ are consistent with that of \cite{Gracey:2015tta,Kompaniets:2021hwg}.


We also computed $Z_g$ and $Z_h$ directly with the help of the OPE,
\begin{equation}
\varphi(x)\bar{\varphi}(0)\sim C_1(x)\sigma(0)+\mathcal{O}(|x|^{-1})~~~,~~~\sigma(x)\sigma(0)\sim C_2(x)\sigma(0)+\mathcal{O}(|x|^{-1})~,
\end{equation}
and the combinations ${Z_g}/{Z_{\sigma}}$ and ${Z_h}/{Z_{\sigma}}$ are determined by the UV finiteness of $C_1$ and $C_2$, respectively as,
\begin{equation}
C_1(x_2)=\frac{Z_g}{Z_{\sigma}}g\tilde{\mu}^{\epsilon}G_1(g_0,h_0;x_0,\del{x}_1,x_2)~~~,~~~
C_2(x_2)=\frac{Z_h}{Z_{\sigma}}h\tilde{\mu}^{\epsilon}G_2(g_0,h_0;x_0,\del{x}_1,x_2)~.
\end{equation}
The corresponding two-point functions are generated by removing one leg from three-point correlation functions $G_1$ and $G_2$ defined as
\begin{equation}
\Bigl\langle \bar{\varphi}(0)\varphi(x_2)\sigma(x_1)\Bigr\rangle
=Z_gg\tilde{\mu}^{\epsilon}G_1(g_0,h_0;x_0,x_1,x_2)~~~,~~~
\Bigl\langle \sigma(0)\sigma(x_2)\sigma(x_1)\Bigr\rangle
=Z_hh\tilde{\mu}^{\epsilon}G_2(g_0,h_0;x_0,x_1,x_2)~.\nonumber
\end{equation}
The results are consistent with previous computation.

The 4-loop correction to the $\gamma_{\varphi^Q}$ is given in Appendix \S\ref{appendix:4-loop-result}.
The scaling dimension of the $\varphi^Q$ at the Wilson-Fisher fixed point is given by
\begin{equation}
\triangle_Q=(2-\epsilon)Q+\gamma_{\varphi^Q}\Bigr|_{g=g^*, h=h^*}~.
\end{equation}
The leading and sub-leading order terms of $\triangle_Q$ in the large $N$ limit were computed in \cite{Jack:2021ziq}, (see also \cite{Vasiliev:1981yc,Vasiliev:1981dg,Vasiliev:1982dc}),
\begin{equation}
\triangle_Q=(2-\epsilon )Q+\frac{Q}{N}\triangle_Q^1+\frac{Q}{N^2}\triangle_Q^2+\mathcal{O}(\frac{1}{N^3})~.
\end{equation}
Our perturbative computation gives
\begin{equation}
\triangle_Q=(2-\epsilon )Q+\frac{Q}{N}\triangle_Q^1+\frac{Q}{N^2}\triangle_Q^2~,
\end{equation}
with
\begin{eqnarray}
\triangle_Q^1&=&\epsilon (8-6Q)+\epsilon ^{2}\left(\frac{-32}{3}+7Q\right)+\epsilon ^{3}\left(-\frac{56}{9}+\frac{11Q}{2}\right)\nonumber\\
&&~~~~~~~~~~~~~~~~~~~~~~~~~~~~~~~~~~~~~+\epsilon^{4}\left[16\zeta_3-\frac{128}{27}+Q\left(\frac{19}{4}-12\zeta_3\right)\right]~,
\end{eqnarray}
and
\begin{eqnarray}
\triangle_Q^2&=&\epsilon (352-264Q)+\epsilon ^{2}\left(\frac{-6272}{3}+1714Q-180Q^{2}\right)\nonumber\\
&&+\epsilon^{3}\left(\frac{16420}{9}+Q\left(\frac{-3743}{3}-864\zeta_3\right)+576\zeta_3+6Q^{2}(31+48\zeta_3)\right)\nonumber\\
&&+\epsilon ^{4}\left[\frac{38630}{27}+\frac{48\pi^{4}}{5}
+1184\zeta(3)+Q\left(\frac{-14930}{9}-\frac{72\pi^{4}}{5}+336\zeta_3\right)\right.\nonumber\\
&&~~~~~~~~~~~~~~~~~~~~~~~~~~~~~~~~~~~~~~~~~~~~~~~~~~~~~~ +Q^{2}\left.\left(334+\frac{24\pi^{4}}{5}-816\zeta_3\right)\right]~.
\end{eqnarray}
These expressions are consistent with the large $N$ computation to the order $\mathcal{O}(\epsilon^4)$.

\section{The OPE approach beyond scalar theories}
\label{sec:OPE-general}

In the previous section, we demonstrated the OPE approach to the anomalous dimensions using the $\phi^Q$ operators as an example, with a particular emphasis on its viability in high-loop computations. But the OPE method can be applied to general cases, for example the operators with Lorentz indices, in the presence of operator mixing, and in general quantum field theory with fermion and gauge fields. In this section, we will show the potential of the OPE method by computing the beta functions and anomalous dimensions of dimension-2 and 3 operators in the Gross-Neveu-Yukawa (GNY) model.

\subsection{The OPE coefficients of general operators}

Before going to the renormalization computation of detailed models, we need to examine the structure of OPE coefficients of general operators. Let us start with a general OPE without operator mixing as,
\begin{equation}\label{eqn:ope1}
\widehat{\mathcal{O}}_1^R(k-p)\widehat{\mathcal{O}}_2^R(p)\sim C(p)\widehat{\mathcal{O}}_3^R(k)+\cdots~
\end{equation}
where an operator with superscript $R$ stands for the renormalized operator. The large momentum expansion is more naturally performed on the bare correlation functions, and the bare operators are related to the renormalized operators by
\begin{equation}\label{eqn:z-define}
\widehat{\mathcal{O}}_i^R=Z_{\widehat{\mathcal{O}}_i}\widehat{\mathcal{O}}_i^{\tiny \rm{bare}}~.
\end{equation}
The OPE of bare operators reads
\begin{equation}\label{eqn:ope-bare1}
\widehat{\mathcal{O}}_1^{\tiny\rm{bare}}(k-p)\widehat{\mathcal{O}}_2^{\tiny\rm{bare}}(p)\sim C_0(p;g_{i0})\widehat{\mathcal{O}}_3^{\tiny\rm{bare}}(k)+\cdots~
\end{equation}
Comparing \eqref{eqn:ope1} and \eqref{eqn:ope-bare1}, we find that the OPE coefficients satisfy
\begin{equation}
C(p)=\frac{Z_{\widehat{\mathcal{O}}_1}Z_{\widehat{\mathcal{O}}_2}}{Z_{\widehat{\mathcal{O}}_3}}C_0(p;g_{i0})~.
\end{equation}
The coefficients $C(p)$ are UV finite, from which we can generate finiteness conditions. The bare OPE coefficients $C_0$ only depend on $p$ and the bare coupling constants $g_{i0}$. They are readily given by the large momentum expansion of bare correlation functions as,
\begin{equation}
C_0(p;g_{i0})=\mathcal{T}_{\widehat{\mathcal{O}}_3}\Bigl\langle \widehat{\mathcal{O}}_1^{\tiny\rm{bare}}(k-p)\widehat{\mathcal{O}}_2^{\tiny\rm{bare}}(p)\phi_0(k_1)\cdots\phi_0(k_n)\Bigr\rangle~,
\end{equation}
in which $n$ is the length of the $\widehat{\mathcal{O}}_3$, and the operation $\mathcal{T}_{\widehat{\mathcal{O}}_3}$ picks up the component of large momentum expansion corresponding to $\widehat{\mathcal{O}}_3$. It gets contribution from the hard graphs after $n$-cutting of large momentum expansion. For example, for length-2 operator $\widehat{\mathcal{O}}_3:=\frac{1}{2}\phi^2$, we get
\begin{equation}
C_0(p;g_{i0})=\mathcal{T}_{\widehat{\mathcal{O}}_3}\Bigl\langle \widehat{\mathcal{O}}_1^{\tiny\rm{bare}}(k-p)\widehat{\mathcal{O}}_2^{\tiny\rm{bare}}(p)\phi_0(k_1)\phi_0(k_2)\Bigr\rangle
=\Bigl\langle \widehat{\mathcal{O}}_1^{\tiny\rm{bare}}(k-p)\widehat{\mathcal{O}}_2^{\tiny\rm{bare}}(p)\phi_0(k_1)\phi_0(k_2)\Bigr\rangle\Bigl|_{k_i=0}~,
\end{equation}
and for $\widehat{\mathcal{O}}_3:=\frac{1}{2}\partial_{\mu}\phi\partial^{\mu}\phi$ we get
\begin{equation}
C_0(p;g_{i0})
=i^2\frac{\partial}{\partial k_1^{\mu}}\frac{\partial}{\partial k_{2\mu}}\Bigl\langle \widehat{\mathcal{O}}_1^{\tiny\rm{bare}}(k-p)\widehat{\mathcal{O}}_2^{\tiny\rm{bare}}(p)\phi_0(k_1)\phi_0(k_2)\Bigr\rangle\Bigl|_{k_i=0}~.
\end{equation}
From the explicit results of $C_0$ and the UV finiteness conditions, the $Z$-factors can be determined.

The OPE of fundamental fields requires extra scrutiny. Following the definition in \eqref{eqn:z-define}, the $Z$-factor of a fundamental field $\phi$ is not $Z_{\phi}$. To avoid confusion, we denote the $Z$-factor by $Z'_{\phi}$, which is defined as,
\begin{equation}
\phi_R=Z'_{\phi}\phi_0=Z'_{\phi}Z_{\phi}^{\frac{1}{2}}\phi~.
\end{equation}
Since $\phi_R=\phi$, we obtain
\begin{equation}
Z'_{\phi}=Z_{\phi}^{-\frac{1}{2}}~.
\end{equation}
Therefore, for the OPE involving a fundamental field, we should have
\begin{equation}
\phi(k-p)\widehat{\mathcal{O}}'^R(p)\sim C(p) \widehat{\mathcal{O}}^R(k)~,
\end{equation}
and the OPE coefficients satisfy
\begin{equation}
C(p)=\frac{Z_{\widehat{\mathcal{O}}'}Z_{\phi}^{-\frac{1}{2}}}{Z_{\widehat{\mathcal{O}}}}C_0(p;g_{i0})~.
\end{equation}

We remark that, in the momentum space one usually computes the amputated correlation functions instead of the full correlation functions. The large momentum expansion of the amputated correlation functions gives {\sl amputated OPE coefficients}, which contain an extra factor of $Z_{\phi}$ for each fundamental field. Thus the {amputated OPE coefficients} satisfy
\begin{equation}
C^{\tiny\rm{amp}}(p)=\frac{Z_{\widehat{\mathcal{O}}'}Z_{\phi}^{\frac{1}{2}}}{Z_{\widehat{\mathcal{O}}}}C^{\tiny\rm{amp}}_{0}(p;g_{i0})~.
\end{equation}
Throughout this paper, the OPE coefficients always mean amputated OPE coefficients unless otherwise specified, and we will ignore the superscript $amp$ in our equations.

In the presence of operator mixing, the OPE reads
\begin{equation}
\phi(k-p)\widehat{\mathcal{O}}'^R_I(p)\sim C_{I\alpha}(p)\widehat{\mathcal{O}}_{\alpha}^R(k)+\cdots~
\end{equation}
and the OPE coefficients satisfy
\begin{equation}
C_{I\alpha}(p)=Z_{\phi}^{\frac{1}{2}}Z_{IJ}C_{0,J\beta}(p; g_{i0})(Z^{-1})_{\beta\alpha}~,
\end{equation}
in which $Z_{IJ}$ and $Z_{\alpha\beta}$ are the $Z$-matrices of $\widehat{\mathcal{O}}'_I$ and $\widehat{\mathcal{O}}_{\alpha}$ respectively,
\begin{equation}
\widehat{\mathcal{O}}'^R_I=Z_{IJ}\widehat{\mathcal{O}}'^{\tiny\rm{bare}}_J~~~,~~~
\widehat{\mathcal{O}}_{\alpha}^R=Z_{\alpha\beta}\widehat{\mathcal{O}}_{\beta}^{\tiny\rm{bare}}~.
\end{equation}
%

\subsection{Renormalization computation of Gross-Neveu-Yukawa model}

Now let us apply the above general discussions to the GNY model. The Lagrangian of GNY model is
\begin{equation}
\mathcal{L}=i\bar{\psi}_i\slashed{\partial}\psi^i+\frac{1}{2}\partial_{\mu}\phi\partial^{\mu}\phi
-g\phi\bar{\psi}_i\psi^i-\frac{\lambda}{4!}\phi^4~.
\end{equation}
We would like to compute the one and two-loop beta functions and anomalous dimensions of the operators with classical dimension-2 and 3 in this model.

The field $Z$-factors $Z_{\psi}$ and $Z_{\phi}$ are still obtained from the UV finiteness of two-point correlations as,
\begin{eqnarray}
&&Z_{\psi}=1+\frac{g^{2}}{2\epsilon }-\frac{g^{4}(5+4N_f)}{8\epsilon ^{2}}+\frac{g^{4}(1+12N_f)}{16\epsilon }~, \\
&&Z_{\phi}=1+\frac{2g^{2}N_f}{\epsilon }-\frac{3g^{4}N_f}{\epsilon ^{2}}
+\frac{60g^{4}N_f-\lambda ^{2}}{24\epsilon }~.
\end{eqnarray}
The $Z$-factor of the dimension-2 operator $\phi^2$ can be obtained from the following OPE,
\begin{equation}
\phi(p)\phi_R^2(k-p)\sim C(p)\phi(k)~,
\end{equation}
and the OPE coefficient $C(p)$ has the form,
\begin{equation}
C(p)=Z_{\phi^2}Z_{\phi}C_{0}(p;g_0,\lambda_0)~,
\end{equation}
in which $C_0$ can be obtained from the large momentum expansion of 3-point bare correlation functions as,
\begin{equation}
C_0(p;g_0,\lambda_0)=\Bigl\langle \phi_0(p)\phi_0^2(k-p)\phi_0(-k)\Bigr\rangle\Bigr|_{k=0}~.
\end{equation}
Unlike the scalar theories, the two-point integrals in the GNY correlation functions contain loop momentum in the numerators, and cannot be evaluated via \verb!HyperlogProcedures!. We evaluate these integrals with the help of IBP-reduction \cite{Smirnov:2008iw,Maierhofer:2017gsa,Wu:2023upw}.

The coefficient $C(p)$ can be expanded into Taylor series of $\ln \frac{p^2}{\mu^2}$ as,
\begin{equation}
C(p)=c_{0}+c_{1}\ln\frac{p^2}{\mu^2}+c_{2}\ln^2\frac{p^2}{\mu^2}+\cdots
\end{equation}
The UV finiteness of $C(p)$ requires all $c_{i}$'s to be UV finite, which not only fix $Z_{\phi^2}Z_{\phi}$ to the current loop order,
\begin{equation}
Z_{\phi^2}Z_{\phi}=1+\frac{\lambda }{2\epsilon }+\frac{\lambda ^{2}-12g^{4}N_f-2g^{2}\lambda N_f}{2\epsilon ^{2}}+\frac{-\lambda ^{2}+8g^{4}N_f+4g^{2}\lambda N_f}{4\epsilon }~,
\end{equation}
but also fix $Z$-factors of $\lambda_0$ and $g_0$ to certain lower-loop orders.

The anomalous dimension of $\phi^2$ is
\begin{equation}
\gamma_{\phi^2}=-\lambda+4g^{2}N_f +\frac{5\lambda ^{2}}{6}
+2g^{4}N_f-4g^{2}\lambda N_f~.
\end{equation}

We compute $g_0$ from the OPE,
\begin{equation}
\phi(p)\psi^{\alpha}(k-p)\sim C'(p)\psi^{\alpha}(k)~,
\end{equation}
and the OPE coefficient has the form,
\begin{equation}
C'(p)=Z_{\psi}Z_{\phi}^{\frac{1}{2}}C'_{0}(p;\lambda_0,g_0)~.
\end{equation}
Since $Z_{\psi}$ and $Z_{\phi}$ are already known, the UV finiteness of $C'(p)$ gives $g_0$ (and some lower-loop corrections of $\lambda_0$ as by-products) as,
\begin{equation}\label{eqn:g0}
g_0=g\tilde{\mu}^{\epsilon}\left(1-\frac{g^{2}(3+2N_f)}{2\epsilon }+\frac{3g^{4}(3+2N_f)^{2}}{8\epsilon ^{2}}
+\frac{-27g^{4}+24g^{2}\lambda +\lambda ^{2}-144g^{4}N_f}{48\epsilon }\right)~.
\end{equation}
Similarly, we compute $\lambda_0$ from the OPE,
\begin{equation}
\phi(p)\phi(k-p)\sim C''(p)\phi_R^2(k)~,
\end{equation}
and the OPE coefficient has the form,
\begin{equation}
C''(p)=\frac{Z_{\phi}^{\frac{1}{2}}Z_{\phi}^{\frac{1}{2}}}{Z_{\phi^2}}
C''_0(p;\lambda_0,g_0)~.
\end{equation}
Hence we obtain
\begin{eqnarray}
\lambda_0&=&\tilde{\mu}^{2\epsilon}\left(
\lambda +\frac{1}{\epsilon }\left(\frac{3\lambda ^{2}}{2}-24g^{4}N_f-4g^{2}\lambda N_f\right)\right.\nonumber\\
&&~~~~~~+\frac{1}{\epsilon ^{2}}\left(\frac{9\lambda ^{3}}{4}-9g^{2}\lambda ^{2}N_f+6g^{4}\lambda N_f\bigl(-5+2N_f\bigr)+24g^{6}N_f\bigl(3+4N_f\bigr)\right)\nonumber\\
&&~~~~~~\left.+\frac{1}{\epsilon }\left(-\frac{17\lambda ^{3}}{12}-96g^{6}N_f+7g^{4}\lambda N_f+3g^{2}\lambda ^{2}N_f\right)\right)~.\label{eqn:lam0}
\end{eqnarray}
Combining \eqref{eqn:g0} and \eqref{eqn:lam0}, we obtain the beta functions,
\begin{eqnarray}
\beta(g)&=&-g^{3}\left(3+2N_f\right)+\left(2g^{3}\lambda +\frac{g\lambda ^{2}}{12}-\frac{3g^{5}}{4}\bigl(3+16N_f\bigr)\right)~,\\
\beta(\lambda)&=&3\lambda ^{2}-48g^{4}N_f-8g^{2}\lambda N_f+\left(-\frac{17\lambda ^{3}}{3}-384g^{6}N_f+28g^{4}\lambda N_f+12g^{2}\lambda ^{2}N_f\right)~.
\end{eqnarray}

Let us present an additional example about the mixing operators. There are two dimension-3 operators which mix to each other, given by,
\begin{equation}
\widehat{\mathcal{O}}'_I=\left(g\bar{\psi}\psi,\lambda\phi^3\right)~.
\end{equation}
The renormalized operators are
\begin{equation}
\widehat{\mathcal{O}}'^R_I=Z_{IJ}\widehat{\mathcal{O}}'^{\tiny\rm{bare}}_I~,
\end{equation}
and the $Z$-matrix can be derived from the following OPE,
\begin{eqnarray}
&&\phi(p)\widehat{\mathcal{O}}'^R_I(k-p)\sim C_I(p)\phi_R^2(k)~,\\
&&\psi^{\alpha}(p)\widehat{\mathcal{O}}'^R_I(k-p)\sim C'_I(p)\psi^{\alpha}(k)~,
\end{eqnarray}
where the OPE coefficients satisfy
\begin{equation}
C_I(p)=\frac{Z_{\phi}^{\frac{1}{2}}}{Z_{\phi^2}}Z_{IJ}C_{0,J}(p;\lambda_0,g_0)~~~,~~~
C'_I(p)=Z_{\psi}Z_{IJ}C'_{0,J}(p;\lambda_0,g_0)~,
\end{equation}
and
\begin{eqnarray}
C_{0,I}(p;\lambda_0,g_0)&=&\Bigl\langle \widehat{\mathcal{O}}'^{\tiny\rm{bare}}_I(-k_1-k_2-p)\phi(p)\phi(k_1)\phi(k_2)\Bigr\rangle\Bigr|_{k_1,k_2=0}~,\\
C'_{0,I}(p;\lambda_0,g_0)&=&\frac{1}{4}\delta_{\alpha}^{\beta}\Bigl\langle \widehat{\mathcal{O}}'^{\tiny\rm{bare}}_I(k-p)\psi^{\alpha}(p)\bar{\psi}_{\beta}(-k)\Bigr\rangle\Bigr|_{k=0}~.
\end{eqnarray}
The anomalous dimension matrix is
\begin{eqnarray}
\gamma_{IJ}&=&-\frac{\partial}{\partial \ln \mu}Z_{IK}(Z^{-1})_{KJ}\nonumber\\
&=&\begin{pmatrix}
-2g^{2}\lambda -\frac{\lambda ^{2}}{12}+2g^{2}N_f+5g^{4}N_f &
\frac{48g^{4}N_f}{\lambda }-24g^{4}N_f+\frac{384g^{6}N_f}{\lambda }  \\
-2g^{2}\lambda & 2g^{2}N_f+\frac{48g^{4}N_f}{\lambda }-\frac{\lambda ^{2}}{12}-19g^{4}N_f+\frac{384g^{6}N_f}{\lambda }
\end{pmatrix}~.
\end{eqnarray}
It can be checked that for $v=(1,1)$, the matrix satisfies
\begin{equation}
v_I\left(\gamma_{IJ}-\frac{1}{2}\gamma_{\phi}\delta_{IJ}\right)=0~,
\end{equation}
which means
\begin{equation}
v_I\widehat{\mathcal{O}}'^{\tiny\rm{bare}}_I=g_0\bar{\psi}_0\psi_0+\frac{\lambda_0}{3!}\phi_0^3
\end{equation}
is an eigen-operator. This is not surprising since the operator is proportional to the $\phi$'s equation of motion up to a total derivative term,
\begin{equation}
\frac{\delta S}{\delta \phi}=-Z_{\phi}^{\frac{1}{2}}\left(\partial^2\phi_0
+g_0\bar{\psi}_{i0}\psi^i_0
+\frac{\lambda_0}{3!}\phi_0^3\right)~.
\end{equation}
%

\section{Conclusion}
\label{sec:conclusion}

In this work, we present a systematic framework for computing anomalous dimensions and beta functions by combining the OPE method with the large momentum expansion technique. The efficacy of this approach is demonstrated through explicit calculations of anomalous dimensions for both elementary fields and composite operators in $O(N)$-symmetric scalar field theories in four and six-dimensions. Through a careful analysis of the leading order contribution of the expansion, we obtain the UV finiteness conditions for $Z$-factors as eqn.(\ref{eqn:ope-ratio}). By requiring the cancelation of $\epsilon$ poles, a system of algebraic equations can be generated that solves $Z$-factors recursively. The advantages of our method can be summarized as follows. Firstly, the UV divergence is computed by means of two-point functions. Compared with the massive bubble method, it is relatively easier to conduct the IBP reduction or evaluate the master integrals when dealing with two-point functions. The computation of master integrals of two-point functions could be performed analytically, and there is no need for the reconstruction from numeric data. Secondly, the evaluation of the UV divergence in this method is in a sense {\sl global}, {\sl i.e.}, the sub-divergences of graphs are automatically cancelled when summing over all loops. In comparison, in $R^\ast$ operation method the subtraction of sub-divergences is a major difficulty. Thirdly, for the scalar theories, one can integrate two-point functions in coordinate space by the graphical function method. The related Maple package \verb!HyperlogProcedures! has been proven to be a very powerful tool in the dimensional regularization scheme to high loops. We also describe the extension of the OPE method for general operators beyond scalar theories, and demonstrate this extension by the two-loop renormalization computations of the Gross-Neveu-Yukawa model.

A straightforward extension of this paper would be the computation of the anomalous dimensions of the $\phi^Q$ operator in cubic scalar theory to five loops. Using the graphical function method, the five-loop beta function and the anomalous dimensions of field and mass have already been computed in \cite{Borinsky:2021jdb}. Additionally, the perturbative computation of the five and six-loop anomalous dimensions of the $\phi^Q$ operator for all $Q$-dependence in quartic scalar theory has been computed \cite{Jin:2022nqq,Bednyakov:2022guj}. We will perform this renormalization computation at five loops for cubic scalar theory using the new OPE method in the forthcoming paper, and generalize the confirmation of equivalence between cubic and quartic scalar theories at fixed points to five loops.

Although scalar theories are the major concern throughout this paper, the idea of expanding a correlation function as a hard two-point function and a soft lower-point correlation function has been shown to be applicable to any theories, including theories with fermions and gauge fields. However, this generalization will introduce non-trivial numerators in the two-point functions. One possible approach to computing the UV divergence of these two-point functions is to generalize the graphical function method to the integrals with numerator structures \cite{Schnetz:2024qqt}. Alternatively, the method can be combined with the IBP method to reduce the hard two-point functions into master integrals with the help of many available packages  \cite{Smirnov:2008iw,Maierhofer:2017gsa,Wu:2023upw}. The UV divergence can be extracted from the massless two-point master integrals, while the five-loop master integrals have been computed in \cite{Georgoudis:2018olj,Georgoudis:2021onj}. Nevertheless, if we want more inputs of the higher-loop master integrals, it might be a viable option to seek assistance from packages such as the AMFlow \cite{Liu:2022chg}. We will continue the research in various general field theories in future.

Another challenging aspect lies in the application of this method to higher dimensional operators that involve derivatives. For example, in \cite{Derkachov:1997ch}, the symmetric traceless operators $\phi_i\partial_{\mu_1\cdots\mu_n}\phi_j$ are considered, and the $\mathcal{O}(\frac{1}{N})$ and $\mathcal{O}(\frac{1}{N^2})$ order scaling dimensions are computed based on conformal field theory, and in \cite{Cao:2021cdt}, the high mass dimension operators and mixing operators are considered by the $R^\ast$ operation method in the effective field theories. Our primary analysis shows the capability of the OPE approach in these problems, and we plan to carry out a perturbative computation of the anomalous dimensions and verify their results in the future work. Moreover, in the presence of the operator mixing, the $Z$-factors take the form of matrices with multiple entries, and the combinations of $Z$-factors in the OPE coefficients are no longer {\sl ratios}. We need to conduct a more elaborate analysis of the OPE in order to extract $Z$-factors.

\section*{Acknowledgments}

We would like to thank Oliver Schnetz for stimulating conversations on the graphical function method and the usage of  HyperlogProcedures package. We would also like to thank Bo Feng and Gang Yang for helpful discussions. RH is supported by the National Natural Science Foundation of China (NSFC) with Grant No.11805102.

\appendix

\section{Sample UV divergence computation of the two-point functions}
\label{appendix:sample_GF}

Let us explain the computation strategy by an example of 3-loop $\varphi^Q\to Q\varphi$ correlation function with $Q=4$. We can generate 7 independent topologies for this correlation function, labeled by $T_1,\ldots,T_7$. Taking one of them for further investigation, say $T_1$, which is sketched below as,
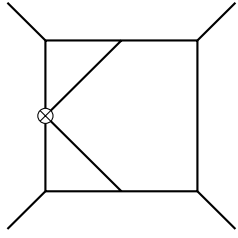
\begin{center}
\begin{tikzpicture}
    \draw [thick] (0.5,-1)--(0.5,1)--(2.5,1)--(2.5,-1)--cycle (1.5,-1)--(0.5,0)--(1.5,1);
    \draw [thick] (0.5,-1)--(0,-1.5) (0.5,1)--(0,1.5) (2.5,1)--(3,1.5) (2.5,-1)--(3,-1.5);
    \draw [fill=white] (0.5, 0) circle [radius=0.1];
    \draw [] (0.43,-0.07)--(0.57,0.07) (0.43,0.07)--(0.57,-0.07);
\end{tikzpicture}
\end{center}
It has four external legs, and if we treat each leg as special, there would be four possible configurations, as highlighted by red lines below
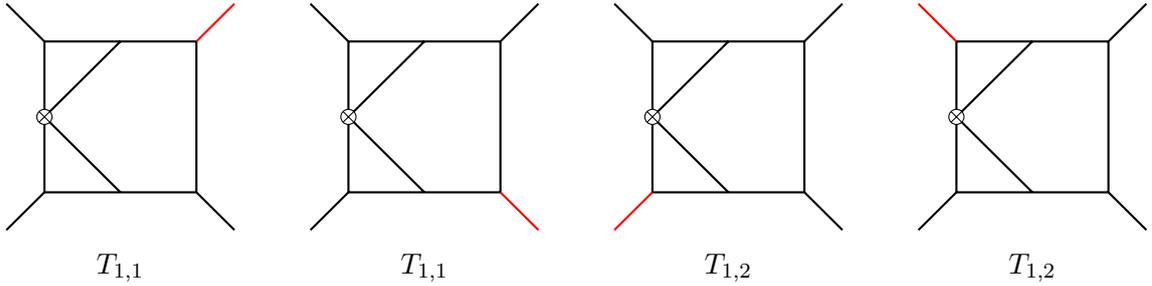
\begin{center}
\begin{tikzpicture}
%
  \node [] at (1.5,0) {
    \begin{tikzpicture}
    \draw [thick] (0.5,-1)--(0.5,1)--(2.5,1)--(2.5,-1)--cycle (1.5,-1)--(0.5,0)--(1.5,1);
    \draw [thick] (0.5,-1)--(0,-1.5) (0.5,1)--(0,1.5) (2.5,-1)--(3,-1.5);
    \draw [thick, red] (2.5,1)--(3,1.5);
    \draw [fill=white] (0.5, 0) circle [radius=0.1];
    \draw [] (0.43,-0.07)--(0.57,0.07) (0.43,0.07)--(0.57,-0.07);
    \end{tikzpicture}
};
  \node [] at (5.5,0) {
    \begin{tikzpicture}
    \draw [thick] (0.5,-1)--(0.5,1)--(2.5,1)--(2.5,-1)--cycle (1.5,-1)--(0.5,0)--(1.5,1);
    \draw [thick] (0.5,-1)--(0,-1.5) (0.5,1)--(0,1.5) (2.5,1)--(3,1.5);
    \draw [thick, red] (2.5,-1)--(3,-1.5);
    \draw [fill=white] (0.5, 0) circle [radius=0.1];
    \draw [] (0.43,-0.07)--(0.57,0.07) (0.43,0.07)--(0.57,-0.07);
    \end{tikzpicture}
};
  \node [] at (9.5,0) {
    \begin{tikzpicture}
    \draw [thick] (0.5,-1)--(0.5,1)--(2.5,1)--(2.5,-1)--cycle (1.5,-1)--(0.5,0)--(1.5,1);
    \draw [thick] (0.5,1)--(0,1.5) (2.5,1)--(3,1.5) (2.5,-1)--(3,-1.5);
    \draw [thick, red] (0.5,-1)--(0,-1.5);
    \draw [fill=white] (0.5, 0) circle [radius=0.1];
    \draw [] (0.43,-0.07)--(0.57,0.07) (0.43,0.07)--(0.57,-0.07);
    \end{tikzpicture}
};
  \node [] at (13.5,0) {
    \begin{tikzpicture}
    \draw [thick] (0.5,-1)--(0.5,1)--(2.5,1)--(2.5,-1)--cycle (1.5,-1)--(0.5,0)--(1.5,1);
    \draw [thick] (0.5,-1)--(0,-1.5) (2.5,1)--(3,1.5) (2.5,-1)--(3,-1.5);
    \draw [thick, red] (0.5,1)--(0,1.5);
    \draw [fill=white] (0.5, 0) circle [radius=0.1];
    \draw [] (0.43,-0.07)--(0.57,0.07) (0.43,0.07)--(0.57,-0.07);
    \end{tikzpicture}
};
  \node [] at (1.5,-2) {$T_{1,1}$};
  \node [] at (5.5,-2) {$T_{1,1}$};
  \node [] at (9.5,-2) {$T_{1,2}$};
  \node [] at (13.5,-2) {$T_{1,2}$};
\end{tikzpicture}
\end{center}
However, because of the symmetry of graphs, for this topology we find that the first and second graphs are isomorphic, while the third and the last graphs are also isomorphic. So computing non-isomorphic graphs we get only two independent graphs, which can be labeled by $T_{1,1}, T_{1,2}$. For 3-loop $Q=4$ case, the 7 independent topologies generate in total 19 independent graphs with one leg special.

Then we should compute the Feynman integrals of these graphs. Since in coordinate space, all external legs but the special one are removed, we are in fact computing simplified Feynman graphs with fewer propagators. For instance, the graph $T_{1,1}$ becomes
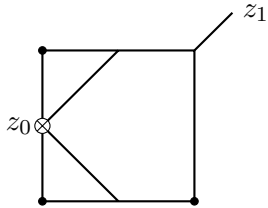
\begin{center}
\begin{tikzpicture}
    \draw [thick] (0.5,-1)--(0.5,1)--(2.5,1)--(2.5,-1)--cycle (1.5,-1)--(0.5,0)--(1.5,1);
    \draw [thick] (2.5,1)--(3,1.5);
    \draw [fill=white] (0.5, 0) circle [radius=0.1];
    \draw [] (0.43,-0.07)--(0.57,0.07) (0.43,0.07)--(0.57,-0.07);
    \node [right] at (3,1.5) {$z_1$};
    \node [left] at (0.5,0) {$z_0$};;
    \draw [fill=black] (0.5,1) circle [radius=0.05] (2.5,-1) circle [radius=0.05] (0.5,-1) circle [radius=0.05];
\end{tikzpicture}
\end{center}
where solid dot in the corner represents the vertex without external leg, which produces double propagators. By the graphical function method, this graph can be computed by the Maple package \verb!HyperlogProcedures!, with command \verb!TwoPoint! by setting $z_0=0$, $z_1=1$ in 6-dimension. It produces a result up to the order $\epsilon$ as\footnote{In $D=6-\epsilon$ convention.},
\begin{equation}
T_{1,1}=(z\bar{z})^{3\epsilon}\left(\frac{1}{3 \epsilon ^3}+\frac{13}{6 \epsilon ^2}+\left(\frac{47}{6}-\frac{\pi ^2}{24}\right)\frac{1}{\epsilon }+\left(\frac{83}{4}-\frac{13 \pi ^2}{48}-\frac{\zeta _3}{12}\right)+\left(\frac{719}{16}-\frac{47 \pi ^2}{48}-\frac{13 \zeta _3}{24}+\frac{13 \pi ^4}{5760}\right) \epsilon \right)~.\nonumber
\end{equation}
Similarly, computing all $T_{i,a}$'s and summing them over by eqn.(\ref{eqn:two-point-sum}), we obtain the corresponding two-point function as $\epsilon$ expansion series in dimensional regularization scheme. We remark that, computation of one degree higher in $\epsilon$ order might already causes drastic increasing in the time consumption. However, for the $\varphi^Q\to Q\varphi$ correlation function in 6-dimensional cubic scalar theory, all graphs are computed to sufficient $\epsilon$ orders in an acceptable short time by a desktop.

\section{The 4-loop correction of the scaling dimension $\Delta_Q$}
\label{appendix:4-loop-result}

Here we present our result of the scaling dimension $\Delta_Q$ in 6-dimensional cubic scalar theory. It has been compared to the large $N$ expansion results in the literatures with agreement. The 4-loop correction of $\Delta_Q$ is given by
\begin{equation}
\Delta_Q^{(4)}=Q\sum_{i=2}^8 g^i h^{8-i}\delta^4_i~,
\end{equation}
in which
\begin{equation}
\delta^4_2=\frac{-69419}{839808}+\frac{29239Q}{559872}+
\left(\frac{35}{1296}-\frac{29Q}{1728}\right)\zeta_3+\left(\frac{1}{6480}-\frac{Q}{8640}\right)\pi^{4}~,
\end{equation}
\begin{equation}
\delta^4_3=\frac{-94597}{186624}+\frac{6665Q}{7776}-\frac{487Q^{2}}{1728}+\left(\frac{97}{144}-\frac{157Q}{144}
+\frac{79Q^{2}}{216}\right)\zeta_3+\left(\frac{-1}{270}+\frac{Q}{240}-\frac{Q^{2}}{1080}\right)\pi^{4}~,
\end{equation}
\begin{equation}
\begin{aligned}
\delta^4_4=&\frac{3960457}{839808}+\frac{719N}{52488}-\frac{1195859Q}{186624}-\frac{17027NQ}{2239488}+\frac{989Q^{2}}{486}-\frac{107Q^{3}}{576}\\
&~~~~~~~~~~~~~~~+\left(\frac{115}{24}-\frac{N}{1296}-\frac{163Q}{27}+\frac{NQ}{1728}+\frac{179Q^{2}}{108}-\frac{Q^{3}}{6}\right)\zeta_3\\
&~~~~~~~~~~~~~~~~~~~~~+\left(\frac{31}{3240}-\frac{43Q}{4320}+\frac{Q^{2}}{540}\right)\pi^{4}
+\left(\frac{-205}{18}+\frac{175Q}{12}-\frac{25Q^{2}}{6}+\frac{5Q^{3}}{12}\right)\zeta_5
\end{aligned}~,
\end{equation}
\begin{equation}
\begin{aligned}
\delta^4_5=&\frac{-88195}{17496}+\frac{63437N}{279936}+\frac{157607Q}{15552}-\frac{739NQ}{2592}-\frac{94337Q^{2}}{15552}+\frac{31NQ^{2}}{486}+\frac{199Q^{3}}{144}-\frac{Q^{4}}{8}\\
&~~~~~~~~~~~~~~+\left(\frac{-635}{36}-\frac{31N}{144}+\frac{6173Q}{216}+\frac{25NQ}{72}-\frac{1381Q^{2}}{108}-\frac{13NQ^{2}}{108}+\frac{19Q^{3}}{12}\right)\zeta_3\\
&
~~~~~~~~~~~~~~~~~~~~~~~~~~~~~~~~~~~~~~+\left(\frac{37}{1080}+\frac{N}{1080}-\frac{11Q}{270}-\frac{NQ}{720}+\frac{11Q^{2}}{1080}+\frac{NQ^{2}}{2160}\right)\pi^{4}\\
&~~~~~~~~~~~~~~~~~~~~~~~~~~~~~~~~~~~~~~~~~~~~~~~~~~~~~~~~~~~~~~+\left(\frac{65}{3}-\frac{75Q}{2}+\frac{115Q^{2}}{6}-\frac{10Q^{3}}{3}\right)\zeta_5
\end{aligned}~,
\end{equation}
\begin{equation}
\begin{aligned}
\delta^4_6=&\frac{13501805}{839808}-\frac{742885N}{279936}-\frac{419N^{2}}{839808}-\frac{593933Q}{34992}+\frac{1535347NQ}{559872}-\frac{29N^{2}Q}{373248}\\
&~~~~~~~~~~~~~~~~~~~~~~~~~~~~~~~~~~~~~~~~~~~~~~+\frac{20201Q^{2}}{7776}-\frac{17365NQ^{2}}{31104}+\frac{197Q^{3}}{288}+\frac{25NQ^{3}}{576}-\frac{5Q^{4}}{16}
\\
&+\left(\frac{19849}{324}+\frac{43N}{18}+\frac{N^{2}}{1296}-\frac{2890Q}{27}-\frac{2633NQ}{864}-\frac{N^{2}Q}{1728}+\frac{12037Q^{2}}{216}+\frac{13NQ^{2}}{16}-\frac{19Q^{3}}{2}\right)\zeta_3\\
&~~~~~~~~~~~~~~~~~~~~~~~~~~~~~~~~~~~~~~~~~~~~~~~~+\left(\frac{1}{2592}-\frac{5N}{648}-\frac{13Q}{2160}+\frac{19NQ}{1728}+\frac{13Q^{2}}{2160}-\frac{NQ^{2}}{288}\right)\pi^{4}\\
&~~~~~~~~~~~~~~~~~~~~~~~~~~~~~~~~~~~~~~~~~~~~~~~~~~~~~~~~~~~~~~~~~+\left(\frac{-1555}{18}+\frac{285Q}{2}-\frac{205Q^{2}}{3}+10Q^{3}\right)\zeta_5
\end{aligned}~,
\end{equation}
\begin{equation}
\begin{aligned}
&\delta^4_7=\frac{-53117}{1944}+\frac{735733N}{139968}+\frac{1129N^{2}}{62208}+\frac{697751Q}{15552}-\frac{578341NQ}{93312}-\frac{1319N^{2}Q}{93312}\\
&~~~~~~~~~~~~~~~~~~~~~~~~~~~~~~~~~~~~~~~~~-\frac{84665Q^{2}}{3888}+\frac{8753NQ^{2}}{5184}+\frac{59N^{2}Q^{2}}{15552}+\frac{613Q^{3}}{144}-\frac{23NQ^{3}}{144}-\frac{7Q^{4}}{16}\\
+&\left(\frac{-3631}{36}-\frac{599N}{216}-\frac{N^{2}}{108}+\frac{9773Q}{54}+\frac{713NQ}{216}+\frac{N^{2}Q}{144}-\frac{3365Q^{2}}{36}-\frac{43NQ^{2}}{36}+\frac{27Q^{3}}{2}+\frac{NQ^{3}}{4}\right)\zeta_3\\
&~~~~~~~~~~~~~~~~~~~~~~~~~~~~~~~~~~~~~~~~~~~~~~~~~~~~~~~~+\left(\frac{41}{1080}+\frac{N}{540}-\frac{53Q}{1080}-\frac{13NQ}{2160}+\frac{Q^{2}}{60}+\frac{NQ^{2}}{360}\right)\pi^{4}
\\
&~~~~~~~~~~~~~~~~~~~~~~~~~~~~~~~~~~~~~~~~~~~~~~~~~~~~~~+\left(140-\frac{5N}{3}-\frac{495Q}{2}+\frac{5NQ}{3}+\frac{255Q^{2}}{2}-20Q^{3}\right)\zeta_5
\end{aligned}~,
\end{equation}
\begin{equation}
\begin{aligned}
\delta^4_8=&\frac{8927569}{419904}-\frac{2504627N}{839808}+\frac{11491N^{2}}{419904}-\frac{N^{3}}{4374}-\frac{7793573Q}{279936}+\frac{428587NQ}{139968}\\
&~~~~~~-\frac{9811N^{2}Q}{559872}
+\frac{19N^{3}Q}{82944}+\frac{3695Q^{2}}{432}-\frac{365NQ^{2}}{2592}-\frac{23N^{2}Q^{2}}{3456}-\frac{7Q^{3}}{24}-\frac{17NQ^{3}}{96}-\frac{21Q^{4}}{64}\\
&+\left(\frac{3067}{72}+\frac{1463N}{324}-\frac{11N^{2}}{216}+\frac{N^{3}}{1296}-\frac{18293Q}{216}-\frac{1001NQ}{216}+\frac{7N^{2}Q}{72}\right.\\
&~~~~~~~~~~~~~~~~~~~~~~~~~~~~~~~~~~~\left.-\frac{N^{3}Q}{1728}+\frac{1855Q^{2}}{36}+\frac{179NQ^{2}}{216}-\frac{17N^{2}Q^{2}}{432}-\frac{77Q^{3}}{8}-\frac{NQ^{3}}{6}\right)\zeta_3\\
&
~~~~~~~~~~~+\left(\frac{41}{1620}-\frac{23N}{12960}+\frac{N^{2}}{2160}-\frac{37Q}{1080}+\frac{29NQ}{4320}-\frac{N^{2}Q}{1440}+\frac{Q^{2}}{90}-\frac{7NQ^{2}}{2160}
+\frac{N^{2}Q^{2}}{4320}\right)\pi^{4}\\
&~~~~~~~~~~~~~~~~~~~+\left(\frac{-640}{9}-\frac{55N}{18}+\frac{775Q}{6}+\frac{35NQ}{12}-70Q^{2}-\frac{5NQ^{2}}{6}+\frac{65Q^{3}}{6}+\frac{5NQ^{3}}{12}\right)\zeta_5
\end{aligned}~.
\end{equation}

\bibliographystyle{JHEP}
\bibliography{Hbib}

\end{document}